\documentclass[%
reprint,
 amsmath,amssymb,
 aps,
prl,
]{revtex4-1}

\usepackage{graphicx}
\usepackage{dcolumn}
\usepackage{bm}
\usepackage{color}
\usepackage{tabularx}
\usepackage{floatrow}
\usepackage[label font=bf,labelformat=simple,caption=false]{subfig}
\usepackage{tikz}
\usepackage{stackengine} 

\newcolumntype{Y}{>{\centering\arraybackslash}X}
\begin{document}

\preprint{APS/123-QED}

\title{Quantum Machine Learning using Gaussian Processes \\ with Performant Quantum Kernels}
  
\author{Matthew Otten}
\email[Correspondence: ]{otten@anl.gov}
\affiliation{%
  Nanoscience and Technology Division, Argonne National Laboratory, Lemont, Illinois, 60439
}%
\author{Im\`{e}ne R. Goumiri}
\author{Benjamin W. Priest}
\author{George F. Chapline}
\author{Michael D. Schneider}
\affiliation{%
  Lawrence Livermore National Laboratory, Livermore, CA, 94450
}%
 
\date{\today}
\author{LLNL-JRNL-808637}
\begin{abstract}
 Quantum computers have the opportunity to be
transformative for a variety of computational tasks. Recently, there have been
proposals to use the unsimulatably of large quantum devices to perform
regression, classification, and other machine learning tasks with quantum
advantage by using kernel methods. While unsimulatably is a necessary condition
for quantum advantage in machine learning, it is not sufficient, as not all
kernels are equally effective. Here, we study
the use of quantum computers to perform the machine learning tasks of
one- and multi-dimensional regression, as well as reinforcement learning, using
Gaussian Processes. By using approximations of performant classical kernels
enhanced with extra quantum resources, we demonstrate that quantum 
devices, both in simulation and on hardware, can perform machine learning tasks
at least as well as, and many times better than, the classical inspiration. Our 
informed kernel design demonstrates a path 
towards effectively utilizing quantum devices for machine learning tasks.
\end{abstract}

\pacs{Valid PACS appear here}
\maketitle

The growing size and quality of quantum computers, especially now that quantum
supremacy has been demonstrated~\cite{Arute:2019aa}, has led to increased interest into practical
applications. Quantum
chemistry~\cite{omalley-prx-2016,peruzzo-ncomms-2014,kandala-nature-2017,colless2018computation}
and quantum  
dynamics~\cite{otten2019noise,smith2019simulating,chiesa2019quantum} have shown
small demonstrations and promising algorithms for the solution of scientific problems. 
Due to the exponential
growth in the Hilbert space with 
the increasing size of quantum computers, machine learning has the potential to be a
strong application of quantum computing, even in the near
term~\cite{biamonte2017quantum}. 
The size of the Hilbert space, however, is not enough to guarantee an efficient
machine learning algorithm. For instance, recent results on the optimization of
quantum neural networks have shown the existence of barren plateaus in the
training landscape~\cite{mcclean2018barren,cerezo2020cost}.
Nevertheless, demonstrations of quantum machine learning include using the
quantum 
computer to encode data in a quantum state (often known as a feature map) and
estimating the inner product, or kernel, of two data points.
The value of the kernel can then be used in classical machine learning tasks,
such as support vector machines for
classification~\cite{havlivcek2019supervised,schuld2019quantum,Bartkiewicz2019ExperimentalKQ}. 
These
demonstrations focused on the general framework of classification using quantum
computers, but did not deeply discuss the important aspect of kernel design.
Other methods have also been studied that directly use the output of the
quantum computer for a regression task~\cite{mitarai2018quantum}.
Though a quantum computer can generally prepare feature maps that are
unsimulatable on classical devices, not every feature map is capable of
providing the same expressibility.
In this manuscript, we demonstrate that a quantum computer can provide
interesting, useful kernels for Gaussian Processes (GPs) for the tasks of regression
and reinforcement learning. We show how classically proven kernels can inspire
new quantum enhanced kernels while still maintaining the important features of
the classical kernel. We start from coherent states, as they approximate the
standard squared exponential kernel~\cite[e.g.,][]{Rasmussen:2005:GPM:1162254},
and show how such states can be prepared on both qudit and qubit hardware,
demonstrating one-dimensional regression using a superconducting qubit
architecture.
Furthermore, we demonstrate that very small quantum devices with only a few
operations can perform higher-dimensional regression better than the squared
exponential kernel by utilizing entanglement and non-classicality. Finally, we
demonstrate that small 
quantum devices can be used in a GP-based reinforcement learning scheme to solve
the classical control problem of driving a one-dimensional car up a hill.


\begin{figure*}
  \begin{center}
    \includegraphics[width=1.0\textwidth]{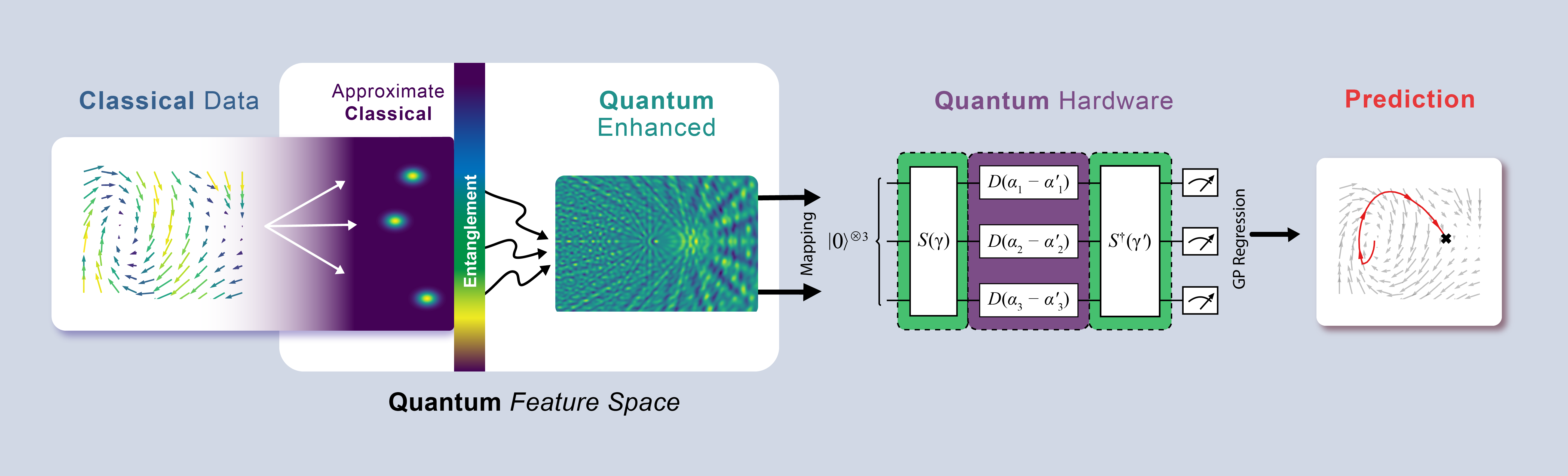}
  \end{center}
  \caption{\label{diagram} 
    Diagram of Gaussian Process (GP) machine learning using quantum enhanced kernels.
    Classical data, here showing a phase space representation of the dynamics of
    a one-dimensional car, is encoded in a quantum state approximating a
    classical feature map. The feature map states for distinct input data dimensions
    can be enhanced using, for example, entanglement.  The overlap of feature
    map states is computed for every pair of data points using quantum hardware. A
    classical computer is then used to perform GP-based machine learning; here,
    predicting an optimal route through phase space.} 
\end{figure*}

\textit{Gaussian Process Regression}
Gaussian Processes are flexible, nonparametric Bayesian models that are performant on a variety of 
statistical and machine learning tasks including regression and
classification~\cite{Neal1997MonteCI,Rasmussen:2005:GPM:1162254}, simulation  
surrogate modeling~\cite{HIGDON20082431}, robotics control~\cite{Ko:2009aa}, and
reinforcement learning~\cite{rasmussen2004}. GP regression, which is summarized
in the Methods section, predicts function 
values at unobserved feature locations by calculation of the kernel between
features, including both observed and 
unobserved. The choice of kernel thus directly influences regression quality.
The kernel between points $\mathbf{x}$ and $\mathbf{x}'$ is implicity
calculated by a nonlinear \textit{feature map}, $\phi(\mathbf{x})$, such that
$k(\mathbf{x},\mathbf{x}') = \langle \phi(\mathbf{x}) | \phi(\mathbf{x'})
\rangle_{\mathcal{F}}$, where the inner product is taken over a reproducing kernel Hilbert 
space $\mathcal{F}$ with kernel $k(\cdot, \cdot)$.
This mapping into higher-dimensional space where observation discrimination becomes
linear is the celebrated ``kernel trick'' applied throughout machine learning.
For quantum computing, we identify the feature Hilbert space $\mathcal{F}$ with the 
Hilbert space of the quantum system~\cite[as in][]{schuld2019quantum,havlivcek2019supervised}.

A quantum computer can represent functions that classical computers cannot efficiently 
calculate~\cite{havlivcek2019supervised}, opening the door for interesting and powerful
kernels exclusive to quantum computers.
However, an unsimulatable feature map is not guaranteed to provide accurate machine 
learning results. Instead of starting with an unsimulatable feature map, we instead start 
with approximations of performant classical feature maps and show that quantum computers 
can enhance their performance by using additional quantum resources, such as entanglement.
Figure~\ref{diagram} demonstrates the general method used in this paper.
Classical data is encoded on the quantum computer using approximations of a
feature map corresponding to a performant classical kernel. 
Extra quantum resources, such as entanglement, lead
to a quantum enhanced feature map, which is mapped to quantum hardware. Through
the quantum device, the kernel matrix is evaluated. Then,
using GPs on classical computers, predictions are made using the calculated
quantum kernel.   

\textit{Coherent State Quantum Kernel}
A general kernel, prepared on a quantum computer, is simply the overlap of two
quantum states, $k(\mathbf{x},\mathbf{x}') = \langle \phi(\mathbf{x}) |
\phi(\mathbf{x'})  \rangle$, where $|\phi(x)\rangle$ is a
quantum circuit or other state preparation
procedure~\cite{schuld2019quantum,havlivcek2019supervised}. There is freedom in
the choice of feature map; different feature maps will have differing performance.
Here, we construct quantum kernels inspired by coherent states because 
(i) the coherent state is easy to prepare on a variety of quantum devices
and, (ii) the inner product of canonical coherent states produces the 
classically popular squared exponential kernel.
A general (canonical) coherent state of complex parameter
$\alpha$ is defined as
\begin{equation}\label{coherent_state}
  |\alpha \rangle = e^{-|\alpha|^2/2} \sum_{n=0}^{\infty} 
    \frac{\alpha^n}{\sqrt{n!}}|n\rangle.
\end{equation}
The equation for the magnitude squared of the overlap between two canonical
coherent states is $ |\langle \alpha | \beta \rangle |^2 = e^{-|\alpha - \beta|^2} $, 
exactly the same form as the squared exponential kernel. 
The input data can be encoded in a variety of
ways, but we choose to encode the data as
$\alpha_i = x_i / (\sqrt{2} c_i)$, where $c_i$ are real hyperparameters, to
maintain consistency with the standard definition of the squared exponential
kernel. We also allow for an additional hyperparameter which scales the result
by a constant, leading to the coherent state kernel
\begin{equation}\label{coherent_kernel}
  k(\mathbf{x},\mathbf{x}') = s \prod_i \Big|\langle  \frac{x_i}{\sqrt{2} c_i}\Big| \frac{x_i'}{\sqrt{2} c_i} \Big\rangle\Big|^2,
\end{equation}
where the product is taken over data dimensions, $s$ is a scaling
hyperparameter, and $| \frac{x_i'}{\sqrt{2} c_i} \rangle$ is a coherent state
with parameter $\alpha = \frac{x_i'}{\sqrt{2} c_i}$. 
  
Although true coherent states can
naturally by prepared on a variety of systems, the ability to provide
non-classicality on these systems can be
difficult~\cite{schumaker1986quantum}. As such, we look at truncations of the
coherent state that are amenable to
preparation on near-term quantum devices such as weakly anharmonic
superconducting resonator cavities or sequences of qubits, which can more easily
prepare non-classical states due to their ability to perform universal quantum
computation. To prepare these states, we note the full coherent state of
eq.~\eqref{coherent_state} can be 
prepared by applying the displacement operator to the vacuum state, i.e.,
$|\alpha \rangle = D(\alpha) | 0 \rangle = e^{\alpha (b^\dagger -  b)} |
0\rangle,$ 
where $b^\dagger$ is a bosonic creation operator and we have now restricted to
real $\alpha$. Generally, $b^\dagger$ is an
infinite dimensional object, but we can truncate the Hilbert space at some
maximum number of levels $N$, giving a representation of the operator
$b^\dagger$ in a finite Hilbert space,
$\tilde{b}^\dagger_N$. A finite-dimensional
coherent state can 
be prepared that approximates the full coherent state by using the finite-dimensional
operators $\tilde{b}^\dagger_N$ in the displacement operator. This finite-dimensional displacement operator,
$\tilde{D}_N(\alpha) = e^{\alpha   (\tilde{b}^\dagger_N - \tilde{b}_N)}$,
is also a natural operator on many quantum  
architectures, such as superconducting microwave
cavities~\cite{hofheinz2008generation}.  By using the finite-dimensional
displacement operator on such systems, we can easily prepare approximations of
the coherent state and follow with further manipulations using universal gates. 

We will denote kernels using
the finite-dimensional coherent state at some level $N$ as \texttt{C-N}.
For qubit based systems,
the finite-dimensional coherent state is not a natural operation. To prepare it, we first
decompose the $N$ level operator into $\log_2(N)$ Pauli operators to get a qubit
Hamiltonian. We then apply a Trotter decomposition~\cite[e.g.,][]{Berry:2007aa} 
to the exponential of the Hamiltonian to
obtain a sequence of gates that are amenable to qubit based devices; details of
this construction can be found in the Supplementary Information. The Trotter
decomposition approximates the exponential of a sum of non-commuting terms as
the product of exponentials of each term. Additional accuracy can be achieved by
repeating the product of the exponentials with reduced coefficients. 
This becomes exact in the limit of
infinite Trotter steps. We will 
denote kernels using the finite-dimensional coherent state of size $N$ prepared on qubits
with a number of Trotter steps $m$ as \texttt{CQ-N-tm}. The use of the
finite-dimensional Hilbert space, especially in the superconducting cavity and
qubit regimes, allows for simple inclusion of non-Gaussian gates.

The finite-dimensional displacement operator, and the qubit
version thereof, define quantum feature maps that are approximations of the
feature map of the classical squared exponential kernel. To measure the overlap
between two points, we use the `echo'
technique~\cite{havlivcek2019supervised,mitarai2018quantum}, where we measure
the magnitude of the overlap between two points $\mathbf{x}$ and $\mathbf{x'}$
by measuring the population of the vacuum after application of the $\tilde{D}_N(\alpha)$
and $\tilde{D}_N^\dagger(\beta)$ to the vacuum, an example of which is shown in
the circuit in Fig.~\ref{diagram}. Since we are utilizing approximations of 
the coherent state, we can instead implement $\tilde{D}_N(\alpha-\beta)|0\rangle
= \tilde{D}_N^\dagger(\beta)\tilde{D}_N(\alpha)|0\rangle$, reducing the total
number of gates by a factor of two. The overlap between every pair of data
points has to be calculated, resulting in the total number of quantum evaluations
of the kernel scaling as the square of the number of data points.

\begin{figure}
  \includegraphics[width=0.97\columnwidth]{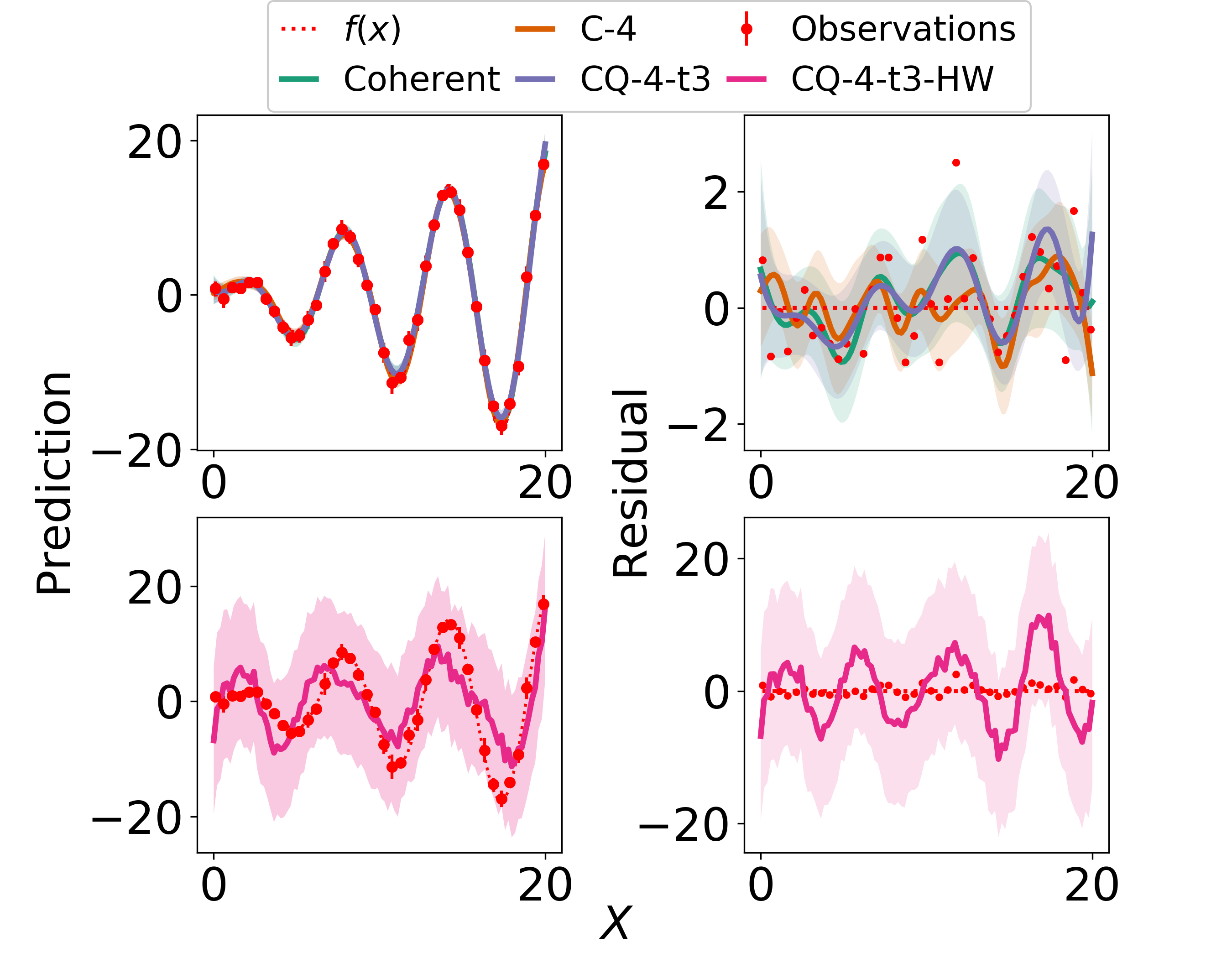}
  \caption{Example of applying quantum kernels to one-dimensional regression of the
    function $f(x)=x\sin(x)$. Acronyms are defined in the text. A small
    truncation with a small number of operations is sufficient to regress well.
    The hardware noise creates a large 
    confidence interval for the \texttt{CQ-4-t3-HW} results, but it still
    captures much of the function.} 
    \label{1dregression} 
\end{figure}

We apply our classically inspired quantum feature maps to one-dimensional
regression of the function $f(x) = x\sin(x)$.
We simulate $M_{tr}=40$ training points uniformly spaced in the region [0.1, 19.9] and  
perturb them with heteroscedastic Gaussian noise $\epsilon_i \sim 
\mathcal{N}(0, \sigma_i^2)$. We assume that the measurement variances 
$\sigma^2_i \sim \mathcal{U}[0, 1]$ are known. 
We then regress on $M_{te}=100$ test points
in the region [0,20], plotting the posterior mean as well as the 95\% confidence interval. 
We use SciKit-Learn~\cite{scikit-learn} to implement the GP regression and
simulated the \texttt{C-N} and \texttt{CQ-N-tm} kernels
using both the QuTiP~\cite{johansson2013qutip} and
QuaC~\cite{QuaC:20,otten-prb-2015} simulation packages. The 
results and residuals of the regression are shown in Fig.~\ref{1dregression}.
The \texttt{C-4} kernel is able to
regress at the same quality as the coherent state, while the two level case
performs much worse (see Supplementary Information). The same
implemented on qubits requires three Trotter steps (\texttt{CQ-4-t3}) to
regress at the same quality as the coherent state; smaller numbers of 
Trotter steps are not able to regress well (see Supplementary Information). This 
demonstrates that a very small quantum device with only a few operations,
whether on a qudit or qubit platform, is able to regress $f(x) = x\sin(x)$ just
as well as the coherent state. In
the Supplementary Information, we apply the same regression technique to other
functions, as well as with other levels of truncation and number of Trotter
steps. The largest Hilbert space necessary was only sixteen states with, for the
qubit kernel, only six Trotter steps. 

We also implement the \texttt{CQ-4-t3} kernel on the IBMQ quantum computer,
Boeblingen, denoted by \texttt{CQ-4-t3-HW}. The use of NISQ hardware prompts us to apply an
alternative GP prior  model, given by
\begin{equation}
\mathbf{y} = \mathbf{f} + \boldsymbol{\epsilon} + \boldsymbol{d}.
\end{equation}
Here, $\mathbf{y}$, $\mathbf{f}$, and $\boldsymbol{\epsilon}$ are as in 
eq.~\eqref{eq:gpmodel} (see Methods), whereas $\mathbf{d}_i \sim \mathcal{N}(0, \sigma^2_d)$ 
are i.i.d. Gaussian variates to capture the additional model discrepancy due to the 
hardware noise~\cite{kennedy2001}.
The noise inherent in the quantum computer results in much larger confidence 
intervals. 
However, note that the addition of model discrepancy term $\mathbf{d}$ 
allows the quantum computer to roughly capture the periodicity of the response function, 
all within the confidence intervals. Furthermore, the model discrepancy term
allows the GP framework to include the effects of the sampling noise and
decoherence from the quantum computer on the regression task. It could also
potentially be used to understand the noise and provide application specific
benchmarking.

Examining the results for various elements of the kernel directly
can also lead to important insights. The diagonal of the kernel should be
identically 1 multiplied by a prefactor; however, even when correcting measurement
errors using Qiskit~\cite{Qiskit}, the average of the diagonal elements is 0.98,
without the prefactor.
These errors are also present for points that are far apart in the
feature space. Because we have the simulated kernel, we can compare small
values of both the simulated and hardware kernels without the prefactor and model
discrepancy terms. For the simulated kernel, training points 4 and 20 are
considered very far away from each other and the kernel for these two points has
a value of 1.7e-4. The 
corresponding point for the hardware kernel has a value of 0.02, significantly
different. For the hardware kernel to have achieved the same value, it would
need to have had only 1 shot out of the total 8192 shots not in the vacuum
state. Various sources of noise, 
such as decoherence and remaining unmitigated readout noise, make the
distinction between very similar and very distinct points difficult. Additional
error mitigation techniques, such as noise
extrapolation~\cite{otten2019recovering,otten2019accounting,kandala2019error,temme2017error,li2017efficient},
could help alleviate these problems, but this still demonstrates an important
restriction on near-term hardware for kernel-based quantum machine learning
methods. Comparing very similar and very distant points is difficult in a noisy,
shot-limited environment. 
Further details of the hardware implementation, including the full Gram matrix, 
as well as the optimal hyperparameters for all kernels, can be found in the
Supplementary Information. 

{\it Entanglement Enhanced Learning}
Up until now, our quantum kernels have been simply approximations of classical
kernels and offered no route to quantum advantage. Quantum advantage could be
obtained by adding additional operations with additional universal gates after
the preparation of finite-dimensional coherent
states~\cite{havlivcek2019supervised,mitarai2018quantum}. Another route, which
we explore here, is adding a hyperparameter that entangles the coherent states
representing different data dimensions. Recall that the default coherent state
(and standard classical squared 
exponential) kernel would simply take the product of the kernels of each data
dimension (see eq.~\eqref{coherent_kernel}). As an enhancement to this, we use
the multi-mode squeezing operator 
to generate entanglement between the data dimensions for a multi-dimensional
regression problem. Squeezing is a standard experimental technique for
generating interesting quantum states, and finds use in quantum metrology, as it
increases the sensing precision in one quadrature at the cost of
others~\cite{loudon1987squeezed,degen2017quantum,kolobov1999spatial}. 
Two-mode squeezing has been demonstrated 
in a variety of
architectures~\cite{patil2015thermomechanical,kurochkin2014distillation,korolkova2002polarization}, 
including superconducting 
cavities~\cite{eichler2011observation}. 
Higher mode squeezing and other multimode interactions, while more difficult,
have also been
demonstrated~\cite{schumaker1987four,yukawa2008experimental}. 

To generate the entanglement enhanced kernel, we first apply the two-mode
squeezing operator to a pair of data dimensions, followed by the displacement
operator, to generate two-mode squeezed coherent states. This is shown in the
circuit in Fig.~\ref{diagram}. The analytical form of
such states and their overlap, in the infinite $N$ limit, is derived in the
Supplementary Information. We use a computationally tractable approximation of
the analytic form to simulate these states. Squeezing represents a purely
non-classical effect that goes beyond 
the approximations of the coherent state. Such squeezing operations can also be 
applied to finite-dimensional coherent states and can further be decomposed into
a qubit form. The amount of squeezing is added as an additional hyperparameter
in the kernel. It is important to note that certain Gaussian states with
quadrature measurements can be
simulated classically~\cite{bartlett2002efficient}; sources of
non-Gaussianity can be added with further control, especially on systems capable
of universal quantum computation.

We apply the two-mode squeezed kernel to the multi-dimensional regression
problem of predicting the dynamics of a car on a hill, given its current
position, $x$; velocity, $v$; and acceleration, $a$.
We entangle each pair of dimensions ($x$ and $v$, $v$ and $a$,
and $x$ and $a$) separately, take the product of each entangled kernel, and then
take the square root of the total kernel to ensure that, in the limit of zero
squeezing, the kernel is equivalent to the unentangled coherent state product
kernel. Further details of the squeezed kernel can be found in the Supplementary
Information.

\begin{table}
  \begin{tabularx}{\columnwidth}{ X | Y | Y | Y | Y }
    \hline
    Kernel & $\overline{R^2_x}$ & $\sigma_x$ & $\overline{R^2_v}$ & $\sigma_v$ \\
    \hline
    Coherent  &  \textbf{0.9985}  &  4.3e-4  &  0.9487  &  0.0277  \\
    Squeezed  &  0.9983  &  5.4e-4  &  \textbf{0.9549}  &  0.0217  \\
    C-8  &  \textbf{0.9985}  &  4.3e-4  &  0.9326  &  0.0496  \\
    C-16  &  \textbf{0.9985}  &  4.3e-4  &  0.9508  &  0.0216  \\
    C-32  &  \textbf{0.9985}  &  4.3e-4  &  0.9487  &  0.0276  \\
  \end{tabularx}
  \caption{Average $R^2$ values and standard deviations for dynamics regression
    using 10 training sets of $M_{tr}=128$ random points each tested on $M_{te}=100$ random test
    points.}\label{ave_r2} 
\end{table}

For our training data, we uniformly sample
$M_{tr}=128$ random positions within the range [$-d,d$], velocities within the range
[$-v,v$], and accelerations within the range $[-v^2/d,v^2/d$]. We then
test the data on $M_{te}=100$ random test points sampled within the same ranges. We
compare the coherent, finite-dimensional coherent, and squeezed kernels over 10
different training and test sets and tabulate the results in Table~\ref{ave_r2}.
The prediction of the next position ($x$) is very good for all of the kernels.
The next velocity ($v$) is not regressed nearly as well, and it is here where
the entanglement provides a quantifiable benefit. 
The average $R^2$ for the coherent kernel is 0.9487, whereas the squeezed has a
significantly higher average score of 0.9549. Interestingly, the
finite-dimensional coherent state of size sixteen also 
performs better than the full coherent state. $R^2$ scores for each training set
can be found in the Supplementary Information.

It is instructive to examine the optimal hyperparameters to understand how much
entanglement is generated for each of the dimensions. A full list of the optimal
hyperparameters, for all kernels, and for all training
sets is given in the Supplementary Information. Notably, in the regression of
the position, $x$, the optimal hyperparameters for every training set include a
large amount of squeezing between the data dimensions representing the current
position and current velocity and include very little squeezing otherwise. This
implies that the current position and current velocity have strong correlation
that can be used to predict the next position. In the regression of the
velocity, $v$, there is, generally, much less squeezing in the optimal
hyperparameters, and the most squeezing is between the current velocity and
current acceleration. The amount of squeezing seen in the optimal
hyperparameters represents the structure of the dynamical equations
which generate the training points, and is supported by a simple linearization
of the underlying equations.

In some training sets, the coherent and squeezed
kernels perform the same for the regression of $v$. This is because the optimal
hyperparameters create very little entanglement between the data dimensions, and
essentially find the best unentangled kernel. This is one of the benefits of
using entanglement enhanced but classically inspired kernels. We have a
guarantee that the squeezed state kernel will perform at least as well as the
coherent state (and, thus, the squared exponential) kernel
in log marginal likelihood because the unentangled coherent 
states are a subset of the squeezed states. This gives a performance guarantee
for the quantum kernels we have outlined in this manuscript; they will perform
at least as well as their classical inspiration. It also allows for an efficient
initial guess of the hyperparameters, which may be important due to the
possibility of barren plateaus in quantum
applications~\cite{mcclean2018barren,cerezo2020cost}.

\begin{figure}
  \centering
  \sidesubfloat[]{\includegraphics[width=0.97\columnwidth]{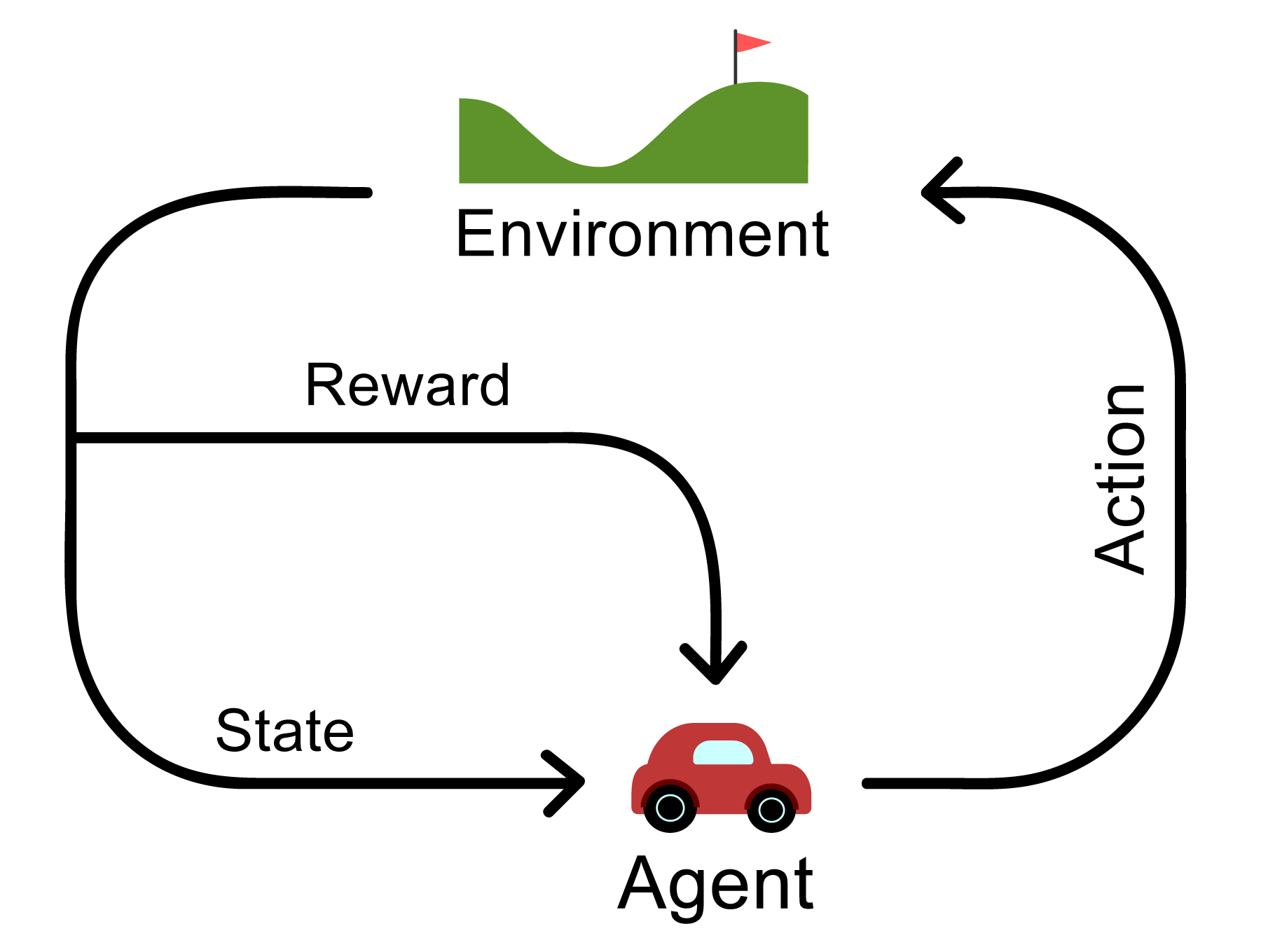}}\\
  \sidesubfloat[]{\includegraphics[width=0.97\columnwidth]{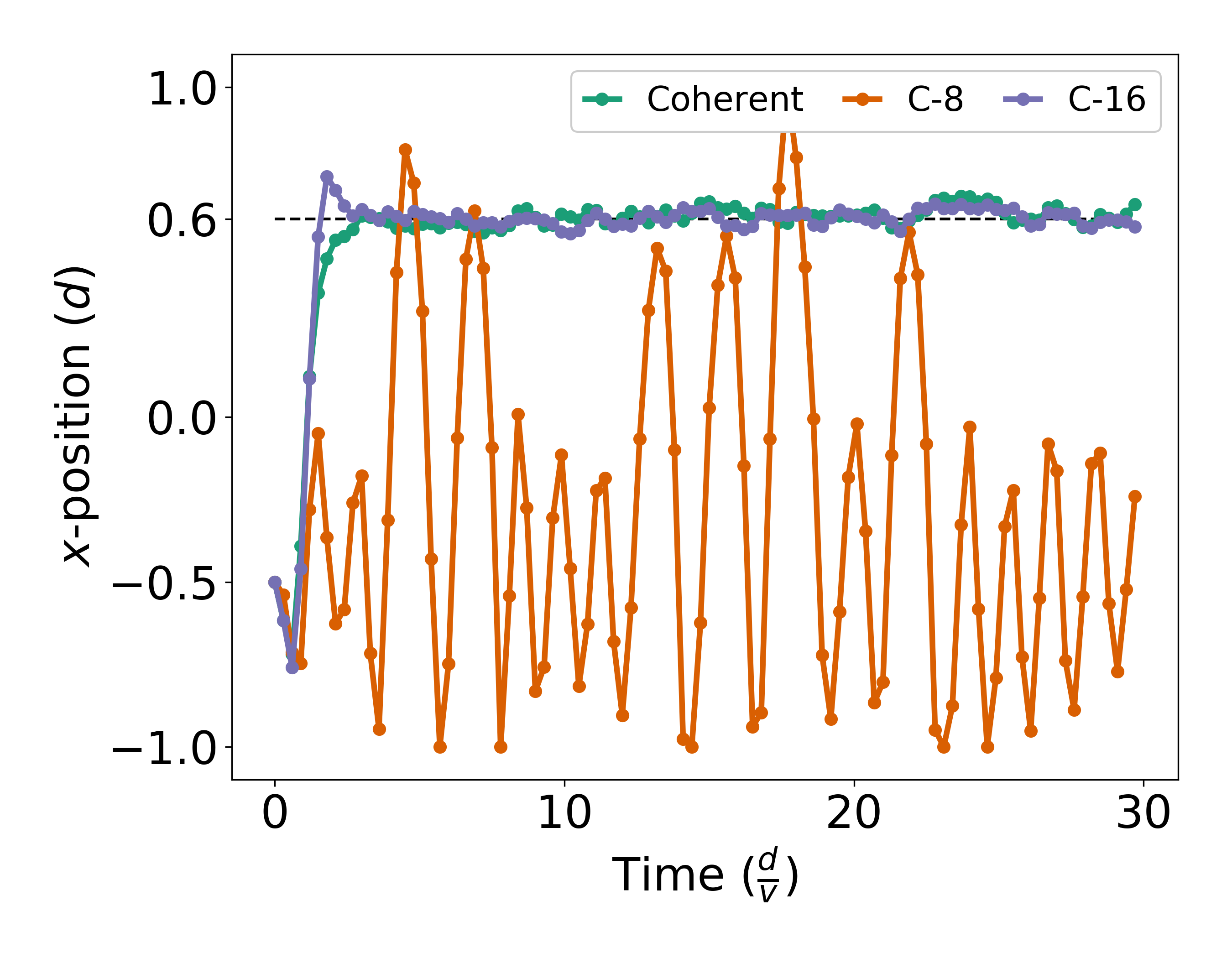}}\\
  \caption{(a) Reinforcement learning framework for the car on a hill problem.
    The goal is to get the car up the hill, to the flag, and keep it there.
    (b) Reinforcement learning results for various quantum kernels. A
    finite-size coherent state of only 16 levels is sufficient to drive the car
    to the goal (black dashed line).} \label{rl_results} 
\end{figure}

{\it Reinforcement Learning}
Given that a small quantum device, with and without entanglement, can accurately
predict the dynamics of a car on a hill, we now explore using quantum kernels to perform
reinforcement learning. In reinforcement learning, an
agent observes its environment, decides a course of action, and gets a reward
based on that action~\cite{kaelbling1996reinforcement}. We use a GP framework to
perform reinforcement learning on 
the classic control problem of driving a car up a
hill~\cite{rasmussen2004,goumiri2020reinforcement}; the agent is
rewarded when the car reaches the goal. GP based reinforcement learning uses
trained GPs to predict the dynamics, and an additional trained GP to predict the
value function. In the GP reinforcement learning process, the car's position
($x$), velocity 
($v$), and acceleration ($a$) are all used in the regression. After
converging the regression of the dynamics and of the value function, we use the
trained GPs to control the car as it moves forward in time. The basic framework
for reinforcement learning and the agent's environment and goal are show in 
Fig.~\ref{rl_results}(a). 

Figure~\ref{rl_results}(b) shows the result of applying GP reinforcement
learning to drive a car up a hill using the quantum kernels described earlier.
The coherent state is able to rapidly reach the goal and stays 
there for the remainder of the simulation time. The finite-dimensional coherent state of
eight levels (\texttt{C-8}), on the other hand, reaches the goal a few times but
cannot keep the car there. With a Hilbert space size of sixteen levels (\texttt{C-16}),
the car again rapidly reaches the goal and stays there. Optimal hyperparameters
for these kernels can be found in the Supplementary Information. This demonstrates that a
small quantum computer can perform reinforcement learning. More demanding
reinforcement learning tasks may necessitate the need for entanglement between
the data dimensions.

{\it Discussion}
We have demonstrated that small, noisy quantum devices with kernels inspired by
and improving on well-studied classical kernels can be performant for a variety of
GP-based machine learning tasks, including one- and 
multi-dimensional regression, as well as reinforcement learning. Utilizing a
GP framework, the method allows for natural inclusion of the inevitable quantum
noise as a model discrepancy term in the GP noise model~\cite[e.g.,][]{kennedy2001}. 
By using informed
kernel design, inspired by performant classical kernels,
we can provide guarantees that, in the absence of noise, the quantum computer
will perform as well as its classical counterpart. The quantum computer,
however, has additional quantum resources not available to the classical kernel,
allowing for the possibility of more performant machine learning. This is
demonstrated by 
using the resource of entanglement, through two-mode squeezing between data
dimensions, in a form unavailable 
to classical kernels. We utilized this quantum enhancement to increase
the performance of multi-dimensional regression of the dynamics of a controlled 
one-dimensional car. 

The enhancement of a standard kernel represents an important element of 
this work to both quantum and classical machine learning. 
The squared exponential kernel, for example, can be described by an
embedding into a quantum feature space through the canonical coherent 
state~\cite{Rasmussen:2005:GPM:1162254,hall_1999}. 
Starting from this embedding, we can examine novel kernel design by examining 
the structure of the embedding. For instance, by using the squared exponential 
kernel as an inspiration for a quantum 
kernel, we start with a feature map that we can analyze, test, and understand
using classical techniques. We can then design feature maps that
have more structure or fill a larger volume of feature space than the canonical
squared exponential feature maps by implementing operations that impose
new structure on the embedding, potentially offering the ability
to learn more complex functions, but without losing the well-studied features of
the classical inspiration.

Though we have demonstrated that quantum devices can prepare interesting feature
maps, the kernel-based GP method we used
still has the draw back that the quantum computer needs calculate the kernel for every
pair of data points. Given the capabilities of
near-term hardware, this could make learning on large datasets infeasible,
especially if hyperparameter optimization becomes costly. Solving the GP problem
directly on the quantum hardware, without having to read off the whole kernel
matrix, is a promising avenue for reducing this overhead. It has already been
reported the HHL algorithm~\cite{harrow2009quantum} could potentially perform
the important matrix inversion step of the GP method~\cite{biamonte2017quantum}, 
but further study is needed to understand 
the full solution of the GP on the quantum device. However, any future algorithm
will still likely require interesting and performant feature maps, such as those
studied here. Other quantum enhanced feature maps based on different classical
kernels, such as the neural network kernel~\cite{buf_paper}, should also be studied to
understand the limits of expressibility on quantum devices. The use of improved
quantum hardware for even more difficult 
machine learning tasks (such as Bayesian search
problems~\cite{chapline2018bayesian} and other problems in filtering,
prediction, and control) represents another promising direction.

\begin{acknowledgments}
  This material is based upon work supported by Laboratory Directed Research and
  Development (LDRD) funding from Argonne National Laboratory, provided by the
  Director, Office of Science, of the U.S. Department of Energy under Contract
  No. DE-AC02-06CH11357. We gratefully acknowledge the computing resources
  provided on Bebop, a high-performance computing cluster
  operated by the Laboratory Computing Resource Center at Argonne National
  Laboratory. Part of this work was performed under the auspices of the
  U.S. Department of Energy by Lawrence Livermore National Laboratory under
  Contract DE-AC52-07NA27344. 
  Funding for this work was provided by LLNL Laboratory Directed 
  Research and Development grant 19-SI-004.
  This research used resources of the Oak Ridge Leadership Computing Facility, which
  is a DOE Office of Science User Facility supported under Contract DE-AC05-
  00OR22725.
\end{acknowledgments}

\textbf{Author Contributions}
All author contributed to the research and writing of the paper. M. O. ran the
GP learning tasks, simulated the quantum kernels, and designed the 
form of the squeezed kernel. I. R. G. provided the GP reinforcement learning
code. B. W. P. provided descriptions and interpretations of GP learning. G. F.
C. and M. D. S. proposed the idea of using coherent states for machine learning tasks. 

\textbf{Data/Code Availability}
The data and code that support the findings of this study are available from the corresponding author on reasonable request.

\textbf{Competing Interests}
The authors declare no competing interests.

\section{Methods}
\subsection{Gaussian Process Regression}
A GP specifies a \emph{prior distribution over functions}, which combined with 
observations specifies a posterior distribution. 
A scalar GP is defined as a distribution $\mathcal{GP}(m(\cdot), k(\cdot, \cdot))$, where
$m(\cdot)$ is a mean function (often set without a loss of generality to zero) and 
$k(\cdot, \cdot)$ is a positive definite reproducing kernel on the input domain.
A GP prior for $n$ possibly multi-variate inputs $X = \{\mathbf{x}_1, \dots,
\mathbf{x}_n\}$ and $n$ scalar outputs $\mathbf{y}$ is 
\begin{align} \label{eq:gpmodel}
\begin{split}
  \mathbf{y} &= \mathbf{f} + \boldsymbol{\epsilon} \\
  \mathbf{f} &\sim \mathcal{N}(\mathbf{0}, K_{X,X}) \\
  \boldsymbol{\epsilon}_i &\sim \mathcal{N}(0, \sigma_i^2).
\end{split}
\end{align}
Here $\mathcal{N}$ is the multivariate Normal (i.e., Gaussian) distribution, $K_{X,X}$ is an 
$n \times n$ matrix whose $(i, j)$th element is $k(\mathbf{x}_i, \mathbf{x}_j)$ and 
$\boldsymbol{\epsilon}$ is heteroscedastic measurement noise with prior variance vector 
$\boldsymbol{\sigma}^2 = [\sigma^2_1, \dots, \sigma^2_n]$.
We assumed throughout that $\boldsymbol{\sigma}^2$ is known a priori.
The joint distribution of $\mathbf{y}$ and the response $\mathbf{f}_*$ of $n_*$ unseen 
data $X_*$ is
\begin{equation}
  \begin{bmatrix} \mathbf{y} \\ \mathbf{f}_*
  \end{bmatrix}
  = \mathcal{N} \left ( 0,
  \begin{bmatrix}
    K_{X,X} + \Sigma & K_{X, X_*} \\
    K_{X_*, X} & K_{X_*, X_*}
  \end{bmatrix} 
      \right ).
\end{equation}
Here, $\Sigma$ is a diagonal matrix whose $(i, i)$th element is $\sigma^2_i$.
This allows us to analytically derive the posterior of the response as 
\begin{align} \label{eq:gp_posterior}
\begin{split}
  \mathbf{f}_* \mid X, X_*, \boldsymbol{\sigma}^2, \mathbf{y} &\sim \mathcal{N}(\bar{\mathbf{f}}_*, C) \\
  \bar{\mathbf{f}}_* &\equiv K_{X_*, X} Q_{X, X}^{-1} \mathbf{y} \\ 
  C &\equiv K_{X_*,X_*} - K_{X_*,X} Q_{X, X}^{-1} K_{X, X_*} \\
  Q_{X, X} &\equiv K_{X,X} + \Sigma. 
\end{split}
\end{align}


\begin{thebibliography}{10}
\expandafter\ifx\csname url\endcsname\relax
  \def\url#1{\texttt{#1}}\fi
\expandafter\ifx\csname urlprefix\endcsname\relax\def\urlprefix{URL }\fi
\providecommand{\bibinfo}[2]{#2}
\providecommand{\eprint}[2][]{\url{#2}}

\bibitem{Arute:2019aa}
\bibinfo{author}{Arute, F.} \emph{et~al.}
\newblock \bibinfo{title}{Quantum supremacy using a programmable
  superconducting processor}.
\newblock \emph{\bibinfo{journal}{Nature}} \textbf{\bibinfo{volume}{574}},
  \bibinfo{pages}{505--510} (\bibinfo{year}{2019}).
\newblock \urlprefix\url{https://doi.org/10.1038/s41586-019-1666-5}.

\bibitem{omalley-prx-2016}
\bibinfo{author}{O`Malley, P.} \emph{et~al.}
\newblock \bibinfo{title}{Scalable quantum simulation of molecular energies}.
\newblock \emph{\bibinfo{journal}{Physical Review X}}
  \textbf{\bibinfo{volume}{6}}, \bibinfo{pages}{031007} (\bibinfo{year}{2016}).

\bibitem{peruzzo-ncomms-2014}
\bibinfo{author}{Peruzzo, A.} \emph{et~al.}
\newblock \bibinfo{title}{A variational eigenvalue solver on a photonic quantum
  processor}.
\newblock \emph{\bibinfo{journal}{Nature communications}}
  \textbf{\bibinfo{volume}{5}}, \bibinfo{pages}{4213} (\bibinfo{year}{2014}).

\bibitem{kandala-nature-2017}
\bibinfo{author}{Kandala, A.} \emph{et~al.}
\newblock \bibinfo{title}{Hardware-efficient variational quantum eigensolver
  for small molecules and quantum magnets}.
\newblock \emph{\bibinfo{journal}{Nature}} \textbf{\bibinfo{volume}{549}},
  \bibinfo{pages}{242} (\bibinfo{year}{2017}).

\bibitem{colless2018computation}
\bibinfo{author}{Colless, J.~I.} \emph{et~al.}
\newblock \bibinfo{title}{Computation of molecular spectra on a quantum
  processor with an error-resilient algorithm}.
\newblock \emph{\bibinfo{journal}{Physical Review X}}
  \textbf{\bibinfo{volume}{8}}, \bibinfo{pages}{011021} (\bibinfo{year}{2018}).

\bibitem{otten2019noise}
\bibinfo{author}{Otten, M.}, \bibinfo{author}{Cortes, C.~L.} \&
  \bibinfo{author}{Gray, S.~K.}
\newblock \bibinfo{title}{Noise-resilient quantum dynamics using
  symmetry-preserving ansatzes}.
\newblock \emph{\bibinfo{journal}{arXiv:1910.06284}}  (\bibinfo{year}{2019}).

\bibitem{smith2019simulating}
\bibinfo{author}{Smith, A.}, \bibinfo{author}{Kim, M.},
  \bibinfo{author}{Pollmann, F.} \& \bibinfo{author}{Knolle, J.}
\newblock \bibinfo{title}{Simulating quantum many-body dynamics on a current
  digital quantum computer}.
\newblock \emph{\bibinfo{journal}{npj Quantum Information}}
  \textbf{\bibinfo{volume}{5}}, \bibinfo{pages}{1--13} (\bibinfo{year}{2019}).

\bibitem{chiesa2019quantum}
\bibinfo{author}{Chiesa, A.} \emph{et~al.}
\newblock \bibinfo{title}{Quantum hardware simulating four-dimensional
  inelastic neutron scattering}.
\newblock \emph{\bibinfo{journal}{Nature Physics}}
  \textbf{\bibinfo{volume}{15}}, \bibinfo{pages}{455--459}
  (\bibinfo{year}{2019}).

\bibitem{biamonte2017quantum}
\bibinfo{author}{Biamonte, J.} \emph{et~al.}
\newblock \bibinfo{title}{Quantum machine learning}.
\newblock \emph{\bibinfo{journal}{Nature}} \textbf{\bibinfo{volume}{549}},
  \bibinfo{pages}{195--202} (\bibinfo{year}{2017}).

\bibitem{mcclean2018barren}
\bibinfo{author}{McClean, J.~R.}, \bibinfo{author}{Boixo, S.},
  \bibinfo{author}{Smelyanskiy, V.~N.}, \bibinfo{author}{Babbush, R.} \&
  \bibinfo{author}{Neven, H.}
\newblock \bibinfo{title}{Barren plateaus in quantum neural network training
  landscapes}.
\newblock \emph{\bibinfo{journal}{Nature communications}}
  \textbf{\bibinfo{volume}{9}}, \bibinfo{pages}{1--6} (\bibinfo{year}{2018}).

\bibitem{cerezo2020cost}
\bibinfo{author}{Cerezo, M.}, \bibinfo{author}{Sone, A.},
  \bibinfo{author}{Volkoff, T.}, \bibinfo{author}{Cincio, L.} \&
  \bibinfo{author}{Coles, P.~J.}
\newblock \bibinfo{title}{Cost-function-dependent barren plateaus in shallow
  quantum neural networks}.
\newblock \emph{\bibinfo{journal}{arXiv preprint arXiv:2001.00550}}
  (\bibinfo{year}{2020}).

\bibitem{havlivcek2019supervised}
\bibinfo{author}{Havli{\v{c}}ek, V.} \emph{et~al.}
\newblock \bibinfo{title}{Supervised learning with quantum-enhanced feature
  spaces}.
\newblock \emph{\bibinfo{journal}{Nature}} \textbf{\bibinfo{volume}{567}},
  \bibinfo{pages}{209--212} (\bibinfo{year}{2019}).

\bibitem{schuld2019quantum}
\bibinfo{author}{Schuld, M.} \& \bibinfo{author}{Killoran, N.}
\newblock \bibinfo{title}{Quantum machine learning in feature hilbert spaces}.
\newblock \emph{\bibinfo{journal}{Physical review letters}}
  \textbf{\bibinfo{volume}{122}}, \bibinfo{pages}{040504}
  (\bibinfo{year}{2019}).

\bibitem{Bartkiewicz2019ExperimentalKQ}
\bibinfo{author}{Bartkiewicz, K.} \emph{et~al.}
\newblock \bibinfo{title}{Experimental kernel-based quantum machine learning in
  finite feature space} (\bibinfo{year}{2019}).

\bibitem{mitarai2018quantum}
\bibinfo{author}{Mitarai, K.}, \bibinfo{author}{Negoro, M.},
  \bibinfo{author}{Kitagawa, M.} \& \bibinfo{author}{Fujii, K.}
\newblock \bibinfo{title}{Quantum circuit learning}.
\newblock \emph{\bibinfo{journal}{Physical Review A}}
  \textbf{\bibinfo{volume}{98}}, \bibinfo{pages}{032309}
  (\bibinfo{year}{2018}).

\bibitem{Rasmussen:2005:GPM:1162254}
\bibinfo{author}{Rasmussen, C.~E.} \& \bibinfo{author}{Williams, C. K.~I.}
\newblock \emph{\bibinfo{title}{Gaussian Processes for Machine Learning
  (Adaptive Computation and Machine Learning)}} (\bibinfo{publisher}{The MIT
  Press}, \bibinfo{year}{2005}).

\bibitem{Neal1997MonteCI}
\bibinfo{author}{Neal, R.~M.}
\newblock \bibinfo{title}{Monte carlo implementation of gaussian process models
  for bayesian regression and classification} (\bibinfo{year}{1997}).

\bibitem{HIGDON20082431}
\bibinfo{author}{Higdon, D.}, \bibinfo{author}{Nakhleh, C.},
  \bibinfo{author}{Gattiker, J.} \& \bibinfo{author}{Williams, B.}
\newblock \bibinfo{title}{A bayesian calibration approach to the thermal
  problem}.
\newblock \emph{\bibinfo{journal}{Computer Methods in Applied Mechanics and
  Engineering}} \textbf{\bibinfo{volume}{197}}, \bibinfo{pages}{2431 -- 2441}
  (\bibinfo{year}{2008}).
\newblock
  \urlprefix\url{http://www.sciencedirect.com/science/article/pii/S0045782507005087}.
\newblock \bibinfo{note}{Validation Challenge Workshop}.

\bibitem{Ko:2009aa}
\bibinfo{author}{Ko, J.} \& \bibinfo{author}{Fox, D.}
\newblock \bibinfo{title}{Gp-bayesfilters: Bayesian filtering using gaussian
  process prediction and observation models}.
\newblock \emph{\bibinfo{journal}{Autonomous Robots}}
  \textbf{\bibinfo{volume}{27}}, \bibinfo{pages}{75--90}
  (\bibinfo{year}{2009}).
\newblock \urlprefix\url{https://doi.org/10.1007/s10514-009-9119-x}.

\bibitem{rasmussen2004}
\bibinfo{author}{Rasmussen, C.~E.} \& \bibinfo{author}{Kuss, M.}
\newblock \bibinfo{title}{Gaussian processes in reinforcement learning}.
\newblock In \emph{\bibinfo{booktitle}{Advances in Neural Information
  Processing Systems}}, vol.~\bibinfo{volume}{16}, \bibinfo{pages}{751--759}
  (\bibinfo{year}{2004}).

\bibitem{schumaker1986quantum}
\bibinfo{author}{Schumaker, B.~L.}
\newblock \bibinfo{title}{Quantum mechanical pure states with gaussian wave
  functions}.
\newblock \emph{\bibinfo{journal}{Physics Reports}}
  \textbf{\bibinfo{volume}{135}}, \bibinfo{pages}{317--408}
  (\bibinfo{year}{1986}).

\bibitem{hofheinz2008generation}
\bibinfo{author}{Hofheinz, M.} \emph{et~al.}
\newblock \bibinfo{title}{Generation of fock states in a superconducting
  quantum circuit}.
\newblock \emph{\bibinfo{journal}{Nature}} \textbf{\bibinfo{volume}{454}},
  \bibinfo{pages}{310--314} (\bibinfo{year}{2008}).

\bibitem{Berry:2007aa}
\bibinfo{author}{Berry, D.~W.}, \bibinfo{author}{Ahokas, G.},
  \bibinfo{author}{Cleve, R.} \& \bibinfo{author}{Sanders, B.~C.}
\newblock \bibinfo{title}{Efficient quantum algorithms for simulating sparse
  hamiltonians}.
\newblock \emph{\bibinfo{journal}{Communications in Mathematical Physics}}
  \textbf{\bibinfo{volume}{270}}, \bibinfo{pages}{359--371}
  (\bibinfo{year}{2007}).
\newblock \urlprefix\url{https://doi.org/10.1007/s00220-006-0150-x}.

\bibitem{scikit-learn}
\bibinfo{author}{Pedregosa, F.} \emph{et~al.}
\newblock \bibinfo{title}{Scikit-learn: Machine learning in {P}ython}.
\newblock \emph{\bibinfo{journal}{Journal of Machine Learning Research}}
  \textbf{\bibinfo{volume}{12}}, \bibinfo{pages}{2825--2830}
  (\bibinfo{year}{2011}).

\bibitem{johansson2013qutip}
\bibinfo{author}{Johansson, J.~R.}, \bibinfo{author}{Nation, P.~D.} \&
  \bibinfo{author}{Nori, F.}
\newblock \bibinfo{title}{Qutip 2: A python framework for the dynamics of open
  quantum systems}.
\newblock \emph{\bibinfo{journal}{Computer Physics Communications}}
  \textbf{\bibinfo{volume}{184}}, \bibinfo{pages}{1234--1240}
  (\bibinfo{year}{2013}).

\bibitem{QuaC:20}
\bibinfo{author}{Otten, M.}
\newblock \bibinfo{title}{Qua{C}: Open quantum systems in {C}, a time-dependent
  open quantum systems solver}.
\newblock \bibinfo{howpublished}{https://github.com/0tt3r/QuaC}
  (\bibinfo{year}{2020}).

\bibitem{otten-prb-2015}
\bibinfo{author}{Otten, M.} \emph{et~al.}
\newblock \bibinfo{title}{Entanglement of two, three, or four plasmonically
  coupled quantum dots}.
\newblock \emph{\bibinfo{journal}{Physical Review B}}
  \textbf{\bibinfo{volume}{92}}, \bibinfo{pages}{125432}
  (\bibinfo{year}{2015}).

\bibitem{kennedy2001}
\bibinfo{author}{Kennedy, M.~C.} \& \bibinfo{author}{O'Hagan, A.}
\newblock \bibinfo{title}{Bayesian calibration of computer models}.
\newblock \emph{\bibinfo{journal}{Journal of the Royal Statistical Society:
  Series B (Statistical Methodology)}} \textbf{\bibinfo{volume}{63}},
  \bibinfo{pages}{425--464} (\bibinfo{year}{2001}).
\newblock
  \urlprefix\url{https://rss.onlinelibrary.wiley.com/doi/abs/10.1111/1467-9868.00294}.
\newblock
  \eprint{https://rss.onlinelibrary.wiley.com/doi/pdf/10.1111/1467-9868.00294}.

\bibitem{Qiskit}
\bibinfo{author}{Abraham, H.} \emph{et~al.}
\newblock \bibinfo{title}{Qiskit: An open-source framework for quantum
  computing} (\bibinfo{year}{2019}).

\bibitem{otten2019recovering}
\bibinfo{author}{Otten, M.} \& \bibinfo{author}{Gray, S.~K.}
\newblock \bibinfo{title}{Recovering noise-free quantum observables}.
\newblock \emph{\bibinfo{journal}{Physical Review A}}
  \textbf{\bibinfo{volume}{99}}, \bibinfo{pages}{012338}
  (\bibinfo{year}{2019}).

\bibitem{otten2019accounting}
\bibinfo{author}{Otten, M.} \& \bibinfo{author}{Gray, S.~K.}
\newblock \bibinfo{title}{Accounting for errors in quantum algorithms via
  individual error reduction}.
\newblock \emph{\bibinfo{journal}{npj Quantum Information}}
  \textbf{\bibinfo{volume}{5}}, \bibinfo{pages}{11} (\bibinfo{year}{2019}).

\bibitem{kandala2019error}
\bibinfo{author}{Kandala, A.} \emph{et~al.}
\newblock \bibinfo{title}{Error mitigation extends the computational reach of a
  noisy quantum processor}.
\newblock \emph{\bibinfo{journal}{Nature}} \textbf{\bibinfo{volume}{567}},
  \bibinfo{pages}{491} (\bibinfo{year}{2019}).

\bibitem{temme2017error}
\bibinfo{author}{Temme, K.}, \bibinfo{author}{Bravyi, S.} \&
  \bibinfo{author}{Gambetta, J.~M.}
\newblock \bibinfo{title}{Error mitigation for short-depth quantum circuits}.
\newblock \emph{\bibinfo{journal}{Physical review letters}}
  \textbf{\bibinfo{volume}{119}}, \bibinfo{pages}{180509}
  (\bibinfo{year}{2017}).

\bibitem{li2017efficient}
\bibinfo{author}{Li, Y.} \& \bibinfo{author}{Benjamin, S.~C.}
\newblock \bibinfo{title}{Efficient variational quantum simulator incorporating
  active error minimization}.
\newblock \emph{\bibinfo{journal}{Physical Review X}}
  \textbf{\bibinfo{volume}{7}}, \bibinfo{pages}{021050} (\bibinfo{year}{2017}).

\bibitem{loudon1987squeezed}
\bibinfo{author}{Loudon, R.} \& \bibinfo{author}{Knight, P.~L.}
\newblock \bibinfo{title}{Squeezed light}.
\newblock \emph{\bibinfo{journal}{Journal of modern optics}}
  \textbf{\bibinfo{volume}{34}}, \bibinfo{pages}{709--759}
  (\bibinfo{year}{1987}).

\bibitem{degen2017quantum}
\bibinfo{author}{Degen, C.~L.}, \bibinfo{author}{Reinhard, F.} \&
  \bibinfo{author}{Cappellaro, P.}
\newblock \bibinfo{title}{Quantum sensing}.
\newblock \emph{\bibinfo{journal}{Reviews of modern physics}}
  \textbf{\bibinfo{volume}{89}}, \bibinfo{pages}{035002}
  (\bibinfo{year}{2017}).

\bibitem{kolobov1999spatial}
\bibinfo{author}{Kolobov, M.~I.}
\newblock \bibinfo{title}{The spatial behavior of nonclassical light}.
\newblock \emph{\bibinfo{journal}{Reviews of Modern Physics}}
  \textbf{\bibinfo{volume}{71}}, \bibinfo{pages}{1539} (\bibinfo{year}{1999}).

\bibitem{patil2015thermomechanical}
\bibinfo{author}{Patil, Y.}, \bibinfo{author}{Chakram, S.},
  \bibinfo{author}{Chang, L.} \& \bibinfo{author}{Vengalattore, M.}
\newblock \bibinfo{title}{Thermomechanical two-mode squeezing in an ultrahigh-q
  membrane resonator}.
\newblock \emph{\bibinfo{journal}{Physical review letters}}
  \textbf{\bibinfo{volume}{115}}, \bibinfo{pages}{017202}
  (\bibinfo{year}{2015}).

\bibitem{kurochkin2014distillation}
\bibinfo{author}{Kurochkin, Y.}, \bibinfo{author}{Prasad, A.~S.} \&
  \bibinfo{author}{Lvovsky, A.}
\newblock \bibinfo{title}{Distillation of the two-mode squeezed state}.
\newblock \emph{\bibinfo{journal}{Physical review letters}}
  \textbf{\bibinfo{volume}{112}}, \bibinfo{pages}{070402}
  (\bibinfo{year}{2014}).

\bibitem{korolkova2002polarization}
\bibinfo{author}{Korolkova, N.}, \bibinfo{author}{Leuchs, G.},
  \bibinfo{author}{Loudon, R.}, \bibinfo{author}{Ralph, T.~C.} \&
  \bibinfo{author}{Silberhorn, C.}
\newblock \bibinfo{title}{Polarization squeezing and continuous-variable
  polarization entanglement}.
\newblock \emph{\bibinfo{journal}{Physical Review A}}
  \textbf{\bibinfo{volume}{65}}, \bibinfo{pages}{052306}
  (\bibinfo{year}{2002}).

\bibitem{eichler2011observation}
\bibinfo{author}{Eichler, C.} \emph{et~al.}
\newblock \bibinfo{title}{Observation of two-mode squeezing in the microwave
  frequency domain}.
\newblock \emph{\bibinfo{journal}{Physical Review Letters}}
  \textbf{\bibinfo{volume}{107}}, \bibinfo{pages}{113601}
  (\bibinfo{year}{2011}).

\bibitem{schumaker1987four}
\bibinfo{author}{Schumaker, B.}, \bibinfo{author}{Perlmutter, S.},
  \bibinfo{author}{Shelby, R.} \& \bibinfo{author}{Levenson, M.}
\newblock \bibinfo{title}{Four-mode squeezing}.
\newblock \emph{\bibinfo{journal}{Physical review letters}}
  \textbf{\bibinfo{volume}{58}}, \bibinfo{pages}{357} (\bibinfo{year}{1987}).

\bibitem{yukawa2008experimental}
\bibinfo{author}{Yukawa, M.}, \bibinfo{author}{Ukai, R.},
  \bibinfo{author}{Van~Loock, P.} \& \bibinfo{author}{Furusawa, A.}
\newblock \bibinfo{title}{Experimental generation of four-mode
  continuous-variable cluster states}.
\newblock \emph{\bibinfo{journal}{Physical Review A}}
  \textbf{\bibinfo{volume}{78}}, \bibinfo{pages}{012301}
  (\bibinfo{year}{2008}).

\bibitem{bartlett2002efficient}
\bibinfo{author}{Bartlett, S.~D.}, \bibinfo{author}{Sanders, B.~C.},
  \bibinfo{author}{Braunstein, S.~L.} \& \bibinfo{author}{Nemoto, K.}
\newblock \bibinfo{title}{Efficient classical simulation of continuous variable
  quantum information processes}.
\newblock \emph{\bibinfo{journal}{Physical Review Letters}}
  \textbf{\bibinfo{volume}{88}}, \bibinfo{pages}{097904}
  (\bibinfo{year}{2002}).

\bibitem{kaelbling1996reinforcement}
\bibinfo{author}{Kaelbling, L.~P.}, \bibinfo{author}{Littman, M.~L.} \&
  \bibinfo{author}{Moore, A.~W.}
\newblock \bibinfo{title}{Reinforcement learning: A survey}.
\newblock \emph{\bibinfo{journal}{Journal of artificial intelligence research}}
  \textbf{\bibinfo{volume}{4}}, \bibinfo{pages}{237--285}
  (\bibinfo{year}{1996}).

\bibitem{goumiri2020reinforcement}
\bibinfo{author}{Goumiri, I.~R.}, \bibinfo{author}{Priest, B.~W.} \&
  \bibinfo{author}{Schneider, M.~D.}
\newblock \bibinfo{title}{Reinforcement learning via gaussian processes with
  neural network dual kernels}.
\newblock \emph{\bibinfo{journal}{arXiv preprint arXiv:2004.05198}}
  (\bibinfo{year}{2020}).

\bibitem{hall_1999}
\bibinfo{author}{{Hall}, B.~C.}
\newblock \bibinfo{title}{{Holomorphic Methods in Mathematical Physics}}.
\newblock \emph{\bibinfo{journal}{arXiv e-prints}}
  \bibinfo{pages}{quant--ph/9912054} (\bibinfo{year}{1999}).
\newblock \eprint{quant-ph/9912054}.

\bibitem{harrow2009quantum}
\bibinfo{author}{Harrow, A.~W.}, \bibinfo{author}{Hassidim, A.} \&
  \bibinfo{author}{Lloyd, S.}
\newblock \bibinfo{title}{Quantum algorithm for linear systems of equations}.
\newblock \emph{\bibinfo{journal}{Physical review letters}}
  \textbf{\bibinfo{volume}{103}}, \bibinfo{pages}{150502}
  (\bibinfo{year}{2009}).

\bibitem{buf_paper}
\bibinfo{author}{{Lee}, J.} \emph{et~al.}
\newblock \bibinfo{title}{{Deep Neural Networks as Gaussian Processes}}.
\newblock \emph{\bibinfo{journal}{ArXiv e-prints}}  (\bibinfo{year}{2017}).
\newblock \eprint{1711.00165}.

\bibitem{chapline2018bayesian}
\bibinfo{author}{Chapline, G.} \& \bibinfo{author}{Otten, M.}
\newblock \bibinfo{title}{Bayesian searches and quantum oscillators}.
\newblock \emph{\bibinfo{journal}{arXiv preprint arXiv:1812.08272}}
  (\bibinfo{year}{2018}).

\end{thebibliography}

%

\end{document}


\preprint{APS/123-QED}

\title{Supplementary Information for Quantum Machine Learning using Gaussian Processes \\ with Performant Quantum Kernels}
  
\author{Matthew Otten}
\email[Correspondence: ]{otten@anl.gov}
\affiliation{%
  Nanoscience and Technology Division, Argonne National Laboratory, Lemont, Illinois, 60439
}%
\author{Im\`{e}ne R. Goumiri}
\author{Benjamin W. Priest}
\author{George F. Chapline}
\author{Michael D. Schneider}
\affiliation{%
  Lawrence Livermore National Laboratory, Livermore, CA, 94450
}%
 
\date{\today}

\pacs{Valid PACS appear here}
\maketitle

\section{Qubit Decomposition of Truncated Displacement Operator}
In this section, we describe the Pauli decomposition of the finite-dimensional
displacement operator, allowing for the preparation of finite-dimensional
approximations of the coherent state on qubit devices. Recall that the
definition of the finite-dimensional creation operator with $N$ levels,
$\tilde{b}_N^\dagger$ is simply the full bosonic creation operator, $b^\dagger$
truncated to a $N\times N$ matrix. For example, given a truncation of $N=4$
levels, we have
\begin{equation}\label{tilde_b}
  \tilde{b}_4^\dagger = 
  \begin{bmatrix}
    0 & 1 & 0 & 0 \\
    0 & 0 & \sqrt{2} & 0 \\
    0 & 0 & 0 & \sqrt{3} \\
    0 & 0 & 0 & 0. \\
  \end{bmatrix}
\end{equation}

The corresponding finite-dimensional displacement operator is simply
$\tilde{D}_N(\alpha) = e^{\alpha   (\tilde{b}^\dagger_N - \tilde{b}_N)}$. To
evaluate this on qubits, we numerically searched for decompositions of
$\tilde{b}_N^\dagger - \tilde{b}_N$ into Pauli strings, resulting in
representations that used the smallest number of qubits possible.
For $N=2$, we have simply $\tilde{b}_2^\dagger - \tilde{b}_2 = Y$.
For $N=4$, we have
\begin{equation}\label{qb4}
  \tilde{b}_4^\dagger - \tilde{b}_4 = \frac{1}{2}\Big((1+\sqrt{3}) YI + (1-\sqrt{3}) YZ + \sqrt{2} (XY - YX)\Big).
\end{equation}
  
For $N=8$, we have
\begin{multline}\label{qb8}
  \tilde{b}_8^\dagger - \tilde{b}_8 = \frac{1}{4}\Big((1+\sqrt{3}+\sqrt{5}+\sqrt{7}) YII + (1+\sqrt{3}-\sqrt{5}-\sqrt{7}) YIZ + (1-\sqrt{3}+\sqrt{5}-\sqrt{7}) YZI + (1-\sqrt{3}-\sqrt{5}+\sqrt{7}) YZZ \\+
  (\sqrt{2}+\sqrt{6}) (XY - YX)I + (\sqrt{2}-\sqrt{7}) (XY - YX)Z + \sqrt{4} (XXY - XYX - YXX + YYY)\Big).
\end{multline}

For $N=16$, we have
\begin{multline}\label{qb16}
  \tilde{b}_{16}^\dagger - \tilde{b}_{16} = \\
  \frac{1}{8}\Big((1+\sqrt{3}+\sqrt{5}+\sqrt{7}+\sqrt{9}+\sqrt{11}+\sqrt{13}+\sqrt{15})
  YIII 
  + (1+\sqrt{3}+\sqrt{5}+\sqrt{7}-\sqrt{9}-\sqrt{11}-\sqrt{13}-\sqrt{15}) YIIZ \\
  + (1+\sqrt{3}-\sqrt{5}-\sqrt{7}+\sqrt{9}+\sqrt{11}-\sqrt{13}-\sqrt{15}) YIZI 
  + (1+\sqrt{3}-\sqrt{5}-\sqrt{7}-\sqrt{9}-\sqrt{11}+\sqrt{13}+\sqrt{15}) YIZZ \\
  + (1-\sqrt{3}+\sqrt{5}-\sqrt{7}+\sqrt{9}-\sqrt{11}+\sqrt{13}-\sqrt{15}) YZII 
  + (1-\sqrt{3}+\sqrt{5}-\sqrt{7}-\sqrt{9}+\sqrt{11}-\sqrt{13}+\sqrt{15}) YZIZ \\
  + (1-\sqrt{3}-\sqrt{5}+\sqrt{7}+\sqrt{9}-\sqrt{11}-\sqrt{13}+\sqrt{15}) YZZI 
  + (1-\sqrt{3}-\sqrt{5}+\sqrt{7}-\sqrt{9}+\sqrt{11}+\sqrt{13}-\sqrt{15}) YZZZ \\
  + (\sqrt{2}+\sqrt{6}+\sqrt{10}+\sqrt{14}) (XY-YX)II
  + (\sqrt{2}+\sqrt{6}-\sqrt{10}-\sqrt{14}) (XY-YX)IZ \\
  + (\sqrt{2}-\sqrt{6}+\sqrt{10}-\sqrt{14}) (XY-YX)ZI
  + (\sqrt{2}-\sqrt{6}-\sqrt{10}+\sqrt{14}) (XY-YX)ZZ\\
  + (\sqrt{4}+\sqrt{12}) (XXY - XYX - YXX + YYY)I + (\sqrt{4}-\sqrt{12}) (XXY -
  XYX - YXX + YYY)Z \\
  + \sqrt{8} (XXXY - XXYX - XYXX - YXXX + XYYY + YXYY + YYXY - YYYX)
\Big).
\end{multline}
This decomposition has obvious structure, but we have not developed a way to
exploit that structure in a scalable manner. Even with a scalable algorithm to
generate the qubit decomposition, it is clear that the resulting sum of Pauli
strings will grow exponentially with the number of qubits, $N_q$, as it includes
all combinations of the identity and Pauli-Z matrices for $N_q-1$ positions. 
For very large
$N$, this would be very inefficient, but, as shown in the main text, the small
truncations show here are sufficient for many interesting machine learning
tasks. Furthermore, since we only need $\log_2(N)$ qubits, a brute force search
may be sufficient. For 
larger $N$, there are scalable methods of preparing coherent states on qubits
using Dicke states~\cite{munoz2009discrete,klco2019systematically}. 

\section{One-Dimensional Regression}
In this section, we provide additional results for the one-dimensional
regression task, including different levels of truncation and number of Trotter
steps for the function presented in the main text, $f(x)=x\sin(x)$, as well
as results for many kernels for two other functions,

\begin{align}
  f_1(x) &= x \sin(\frac{0.65 x}{(1+0.1x)}) \cos(\sin(x)) \label{f1} \\
  f_2(x) &= \frac{0.65 x}{1+0.1x} \label{f2}.
\end{align}

\begin{figure}
  \centering
  \sidesubfloat[]{\includegraphics[width=0.47\columnwidth]{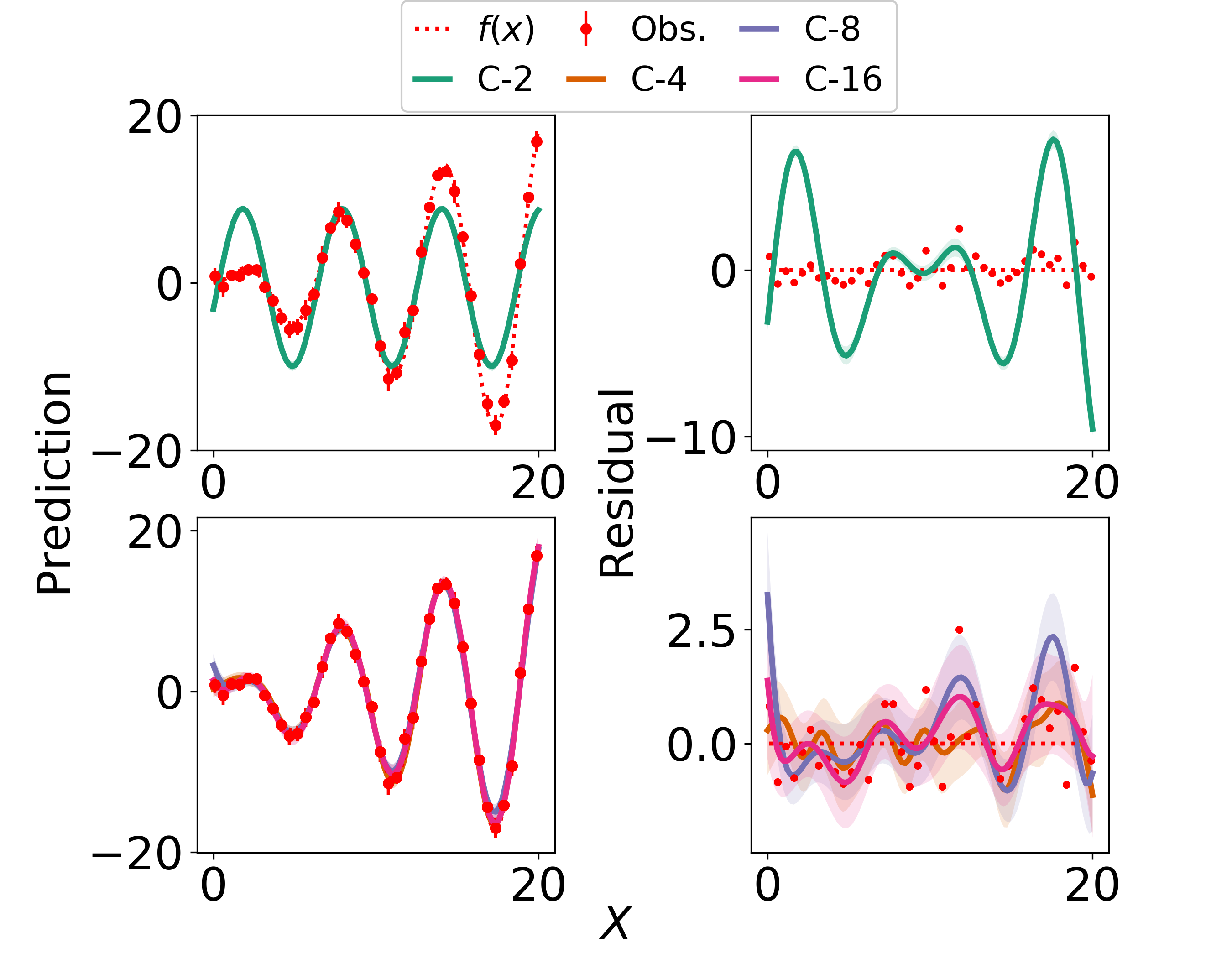}}
  \sidesubfloat[]{\includegraphics[width=0.47\columnwidth]{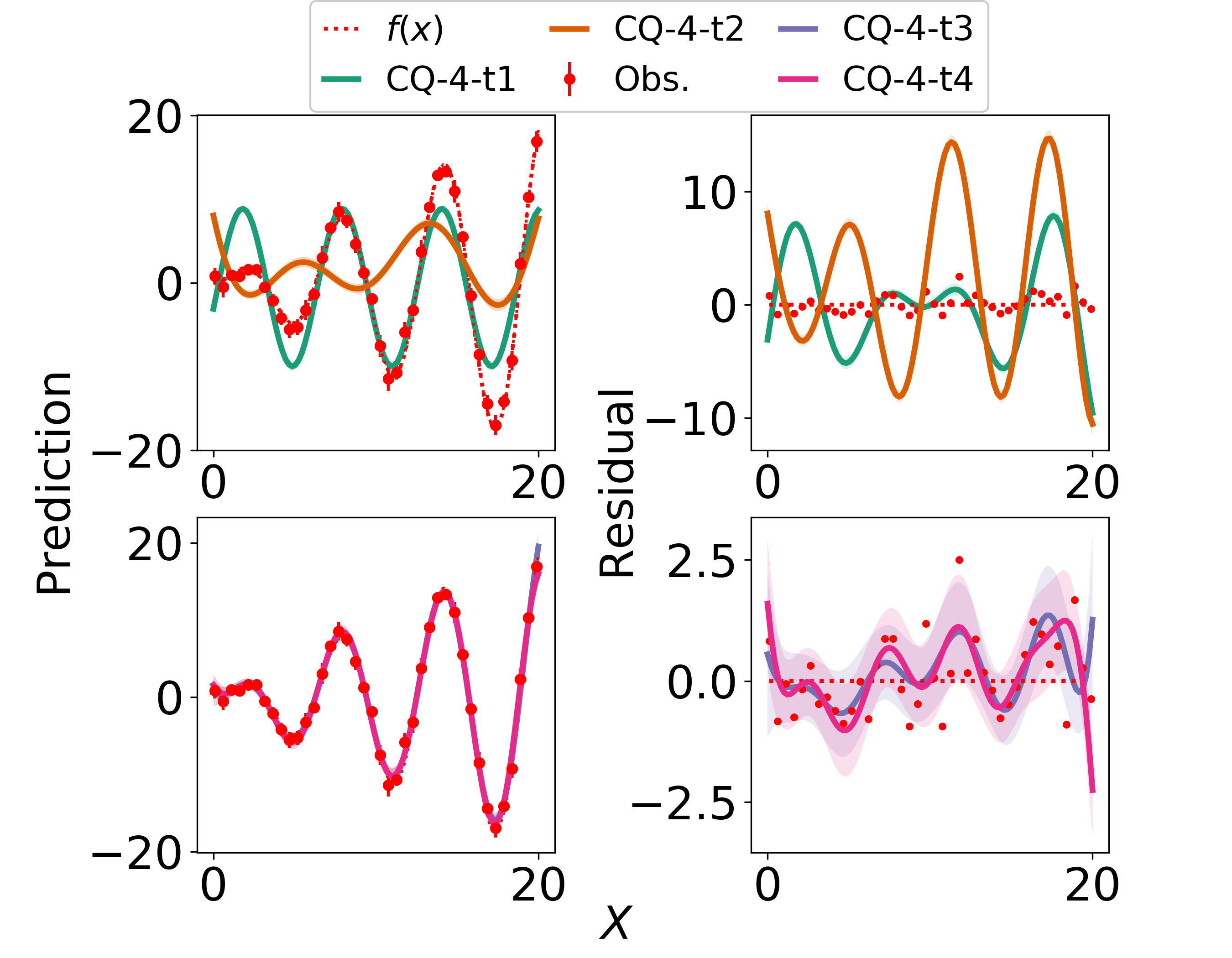}}
  \caption{GP regression of $f(x)=x\sin(x)$. (a) Various levels of truncation
    for the finite-dimensional coherent kernel. Two levels is not sufficient to
    regress this function, but four, eight, and sixteen levels obtain accurate
    regressions. (b) Various numbers of Trotter steps for the finite-dimensional
    coherent kernel of four levels implemented on qubits. One or two Trotter steps
    is not sufficient to regress this function, but three and four Trotter steps
    regress well.} \label{dynamics_regression0} 
\end{figure}

\begin{figure}
  \centering
  \includegraphics[width=0.47\columnwidth]{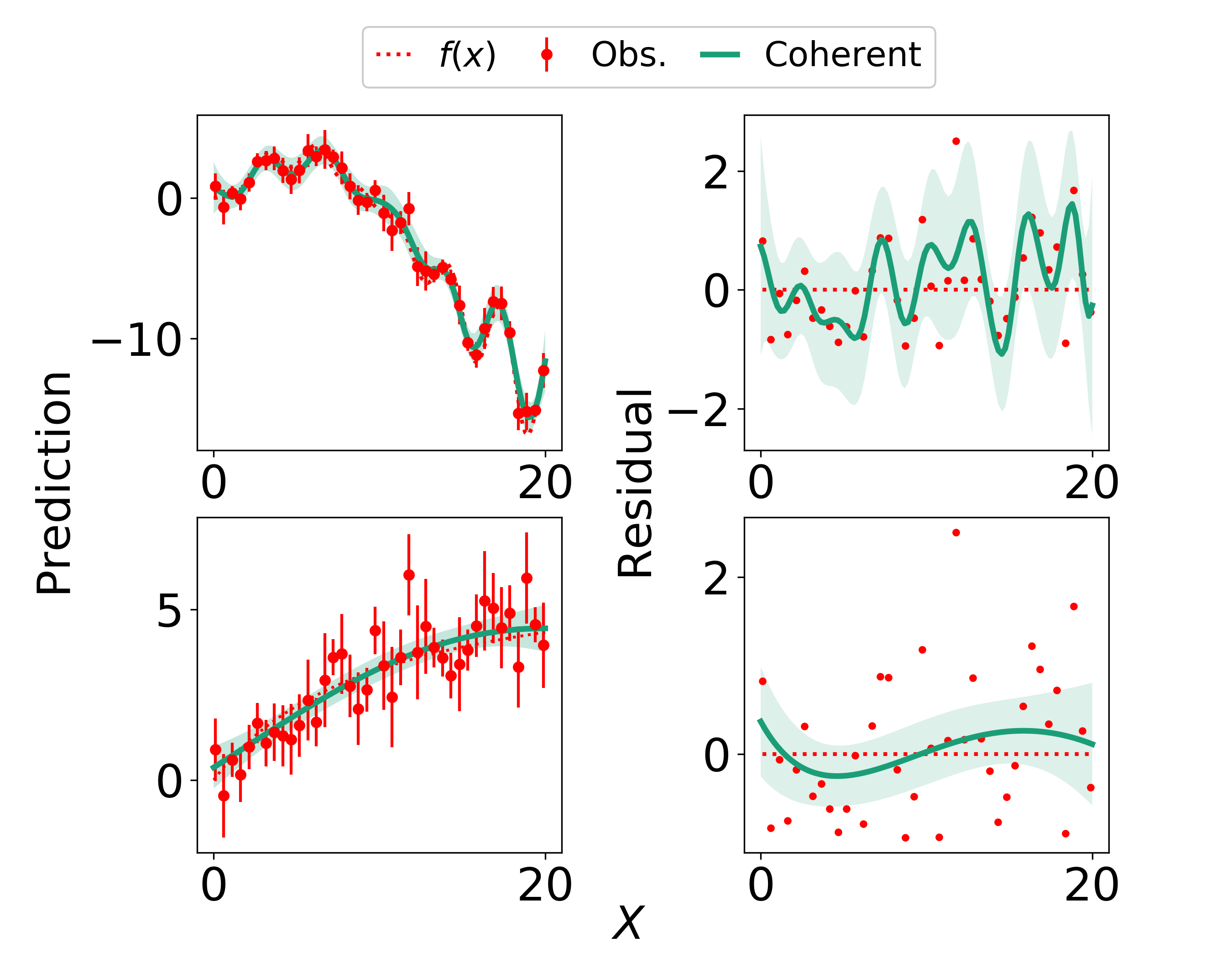}
  \caption{GP regression of $f_1(x)$ (top, eq.~\eqref{f1}) and $f_2(x)$
    (bottom, eq.~\eqref{f2}) using the coherent kernel.} \label{dynamics_regression_1_2} 
\end{figure}

\begin{figure}
  \centering
  \sidesubfloat[]{\includegraphics[width=0.47\columnwidth]{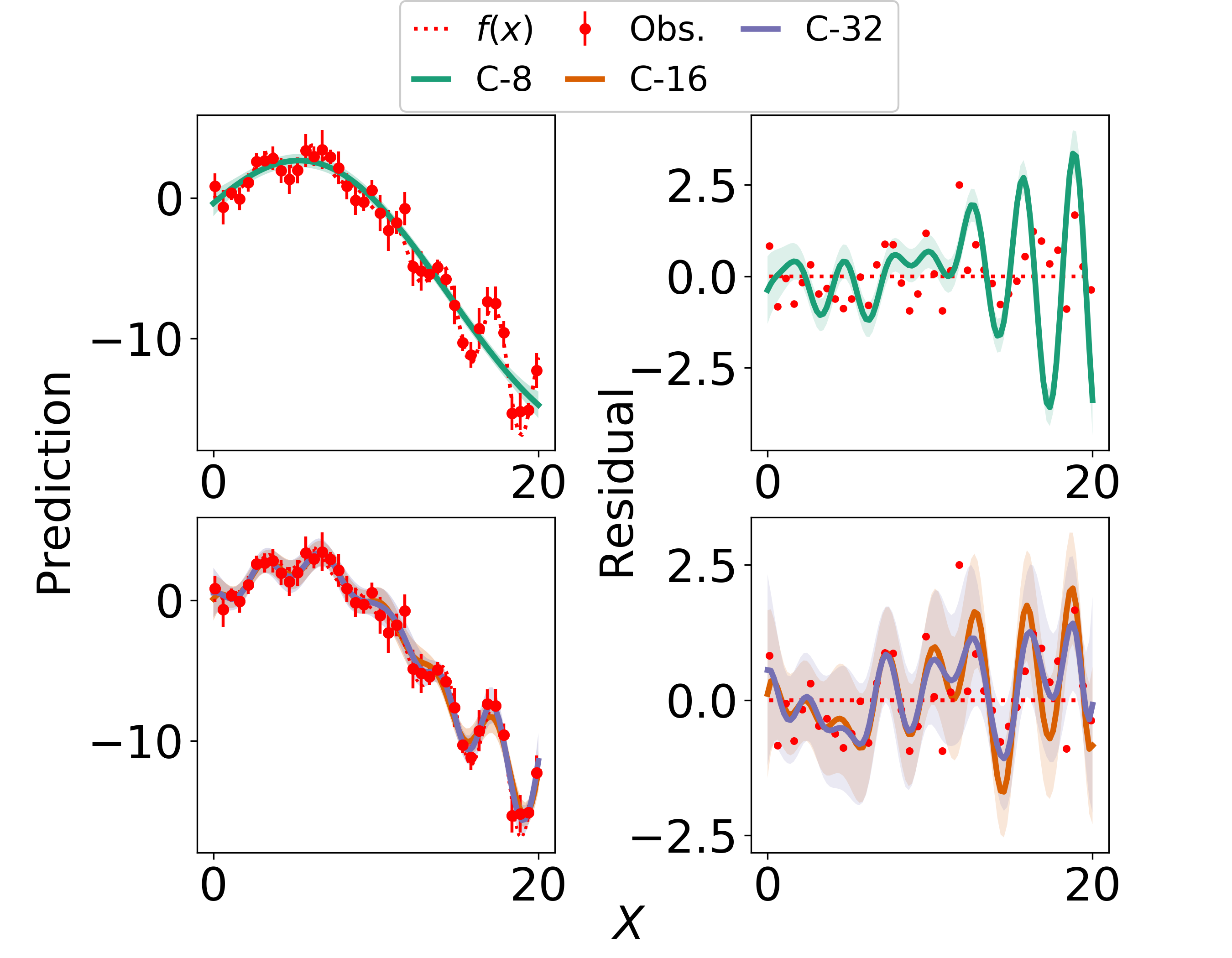}}
  \sidesubfloat[]{\includegraphics[width=0.47\columnwidth]{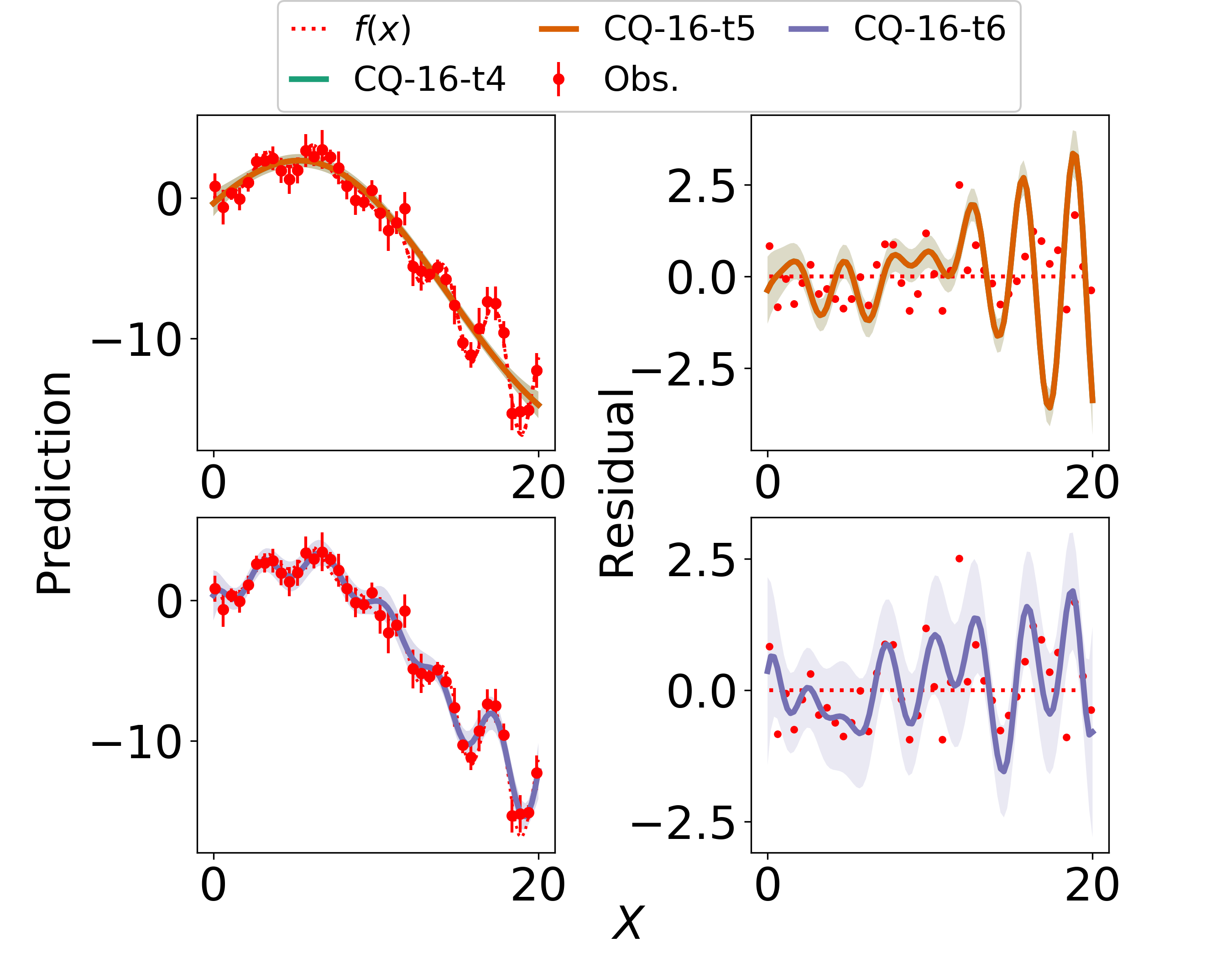}}
  \caption{GP regression of $f_1(x)$ (eq.~\eqref{f1}). (a) Various levels of truncation
    for the finite-dimensional coherent kernel. Eight levels is not sufficient to
    regress this function, but sixteen and thirty-two levels obtain accurate
    regressions. (b) Various numbers of Trotter steps for the finite-dimensional
    coherent kernel of sixteen levels implemented on qubits. Four or five Trotter steps
    is not sufficient to regress this function, but six Trotter steps
    regresses well.} \label{dynamics_regression_1} 
\end{figure}

\begin{figure}
  \centering
  \includegraphics[width=0.47\columnwidth]{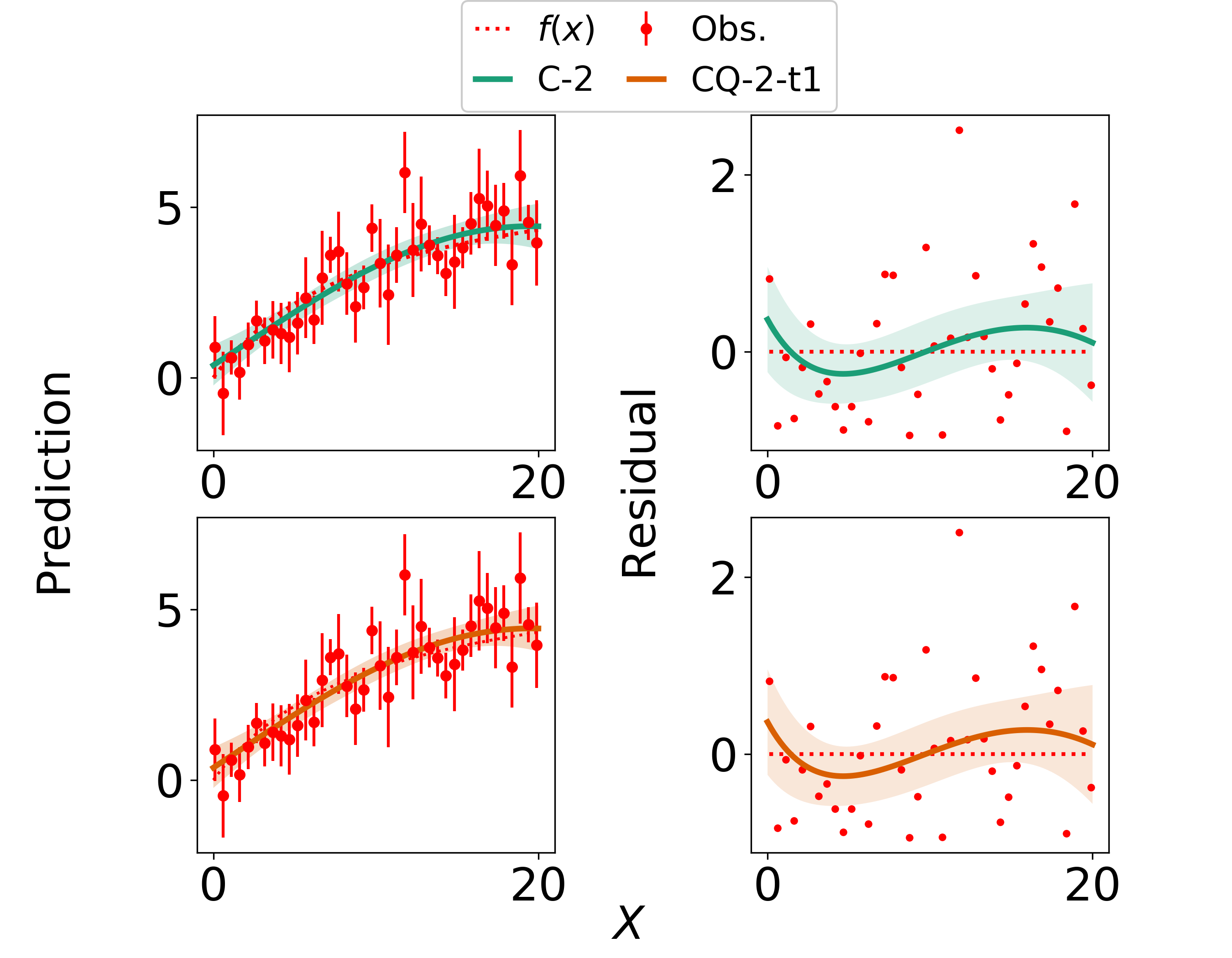}
  \caption{GP regression of $f_2(x)$ (eq.~\eqref{f2}). A truncation of only two
    levels is sufficient to regress this function; since the same implemented on
  qubits requires only one operation ($Y$), a single Trotter step is sufficient.} \label{dynamics_regression_2} 
\end{figure}

In summary, Fig.~\ref{dynamics_regression0} demonstrates the convergence of the
number of levels and number of Trotter steps for the regression problem studied
in the main text. Figure~\ref{dynamics_regression_1_2} shows regression using
coherent kernel for eqs.~\eqref{f1} and \eqref{f2}, providing the benchmark
results for the other kernels. Figure~\ref{dynamics_regression_1} demonstrates
that sixteen levels and five Trotter steps are necessary to regress the harder
function of eq.~\eqref{f1}, while Fig.~\ref{dynamics_regression_2} demonstrates
that $f_2$ can be regressed with only a single qubit.

\clearpage
\subsection{Optimal Hyperparameters for One-Dimensional Regression}
\begin{table*}[!htb]
  \begin{tabular}{ c |c |c  }
    \hline
    Kernel & $s$ & $c_1$ \\
    \hline
    Coherent & 1.000e+02 & 1.764e+00 \\
    \texttt{C-2} & 5.943e+01 & 1.379e+00 \\
    \texttt{C-4} & 1.000e+02 & 1.085e+00 \\
    \texttt{C-8} & 1.000e+02 & 2.921e+00 \\
    \texttt{C-16} & 1.000e+02 & 2.040e+00 \\
    \texttt{CQ-4-t1} & 5.873e+01 & 1.379e+00 \\
    \texttt{CQ-4-t2} & 1.885e+01 & 3.700e+00 \\
    \texttt{CQ-4-t3} & 9.570e+01 & 2.225e+00 \\
    \texttt{CQ-4-t4} & 6.982e+01 & 2.029e+00 \\
  \end{tabular}
  \caption{Optimal hyperparameters for one-dimensional regression of
    $f(x)=x\sin(x)$. The bounds for $s$ were [1e-2,1e2] and the bounds for
    $c_1$ were [1e-3,1e3].}
\end{table*}

\begin{table*}[!htb]
  \begin{tabular}{ c |c |c  }
    \hline
    Kernel & $s$ & $c_1$ \\
    \hline
    Coherent & 3.074e+01 & 1.384e+00 \\
    \texttt{C-8} & 1.000e+02 & 1.076e+01 \\
    \texttt{C-16} & 1.000e+02 & 1.926e+00 \\
    \texttt{C-32} & 3.115e+01 & 1.382e+00 \\
    \texttt{CQ-16-t4} & 1.000e+02 & 1.073e+01 \\
    \texttt{CQ-16-t5} & 1.000e+02 & 1.074e+01 \\
    \texttt{CQ-16-t6} & 9.289e+01 & 1.772e+00 \\
  \end{tabular}
  \caption{Optimal hyperparameters for one-dimensional regression of
    $f_1(x)$, eq.~\eqref{f1}. The bounds for $s$ were [1e-2,1e2] and the bounds for
    $c_1$ were [1e-3,1e3].}
\end{table*}

\begin{table*}[!htb]
  \begin{tabular}{ c |c |c  }
    \hline
    Kernel & $s$ & $c_1$ \\
    \hline
    Coherent & 1.189e+01 & 1.787e+01 \\
    \texttt{C-2} & 1.028e+01 & 1.621e+01 \\
    \texttt{CQ-2-t1} & 1.094e+01 & 1.642e+01 \\
  \end{tabular}
  \caption{Optimal hyperparameters for one-dimensional regression of
    $f_2(x)$, eq.~\eqref{f2}. The bounds for $s$ were [1e-2,1e2] and the bounds for
    $c_1$ were [1e-3,1e3].}
\end{table*}

\section{Additional Hardware Kernel Details}
\begin{figure}
  \centering
  \includegraphics[width=0.47\columnwidth]{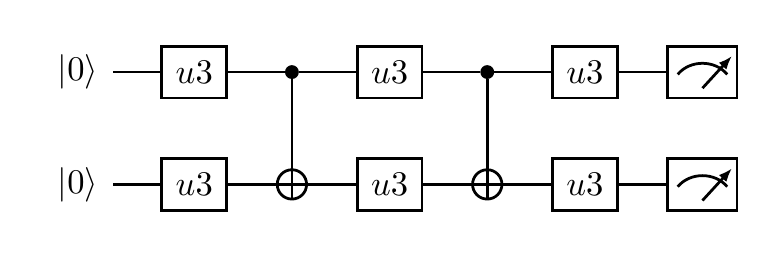}
  \caption{Form of the fully compiled two-qubit circuit used for the
    \texttt{CQ-4-t3-HW} kernel. Each $u3$ gate has different parameters, representing
    different single qubit rotations.} \label{hw_circuit} 
\end{figure}

In this section we provide further details of the use of the IBM quantum
computer, Boeblingen, for GP regression of the function $f(x)=x\sin(x)$. We use
the qubit decomposition of eq.~\eqref{qb4} and use a Trotter
decomposition~\cite{Berry:2007aa} of 
three steps to generate the finite-dimensional coherent kernel on qubits, labeled
\texttt{CQ-4-t3-HW}. Because optimization on the hardware is time consuming, we
instead use the optimal $c_i$ parameters from the \texttt{C-4} kernel as fixed
parameters for the \texttt{CQ-4-t3-HW} kernel. We can still vary other
hyperparameters, such as the scaling prefactor. We then compile the circuits, for
all pairs of data points, using the Qiskit
compiler~\cite{Qiskit}, targeting the Boeblingen backend and using the highest
optimization level. The form of the circuit, for any two points $x$ and $x'$, is
shown in Fig.~\ref{hw_circuit}. We have omitted the parameters of each $u3$
gate; they depend on the specific values of $x$ and the fixed $c_i$. For the
points $x=x'$, we know, a priori, that the resulting circuit should be identity.
However, to keep the amount of noise between each pair of points the same, we
force the evaluation of the kernel for $x=x'$ to have the same form as the
$x\ne x'$ points. We use the full state measurement error mitigation available in
Qiskit Ignis~\cite{Qiskit} to mitigate the effects of measurement error and use 8192 shots for
each circuit. We also symmetrize
the resulting kernel matrix, since, due to various noise sources, the evaluated
values of $k(x,x')$ and $k(x',x)$ are not likely to be equal.

\begin{figure}
  \centering
  \sidesubfloat[]{\includegraphics[width=0.33\columnwidth]{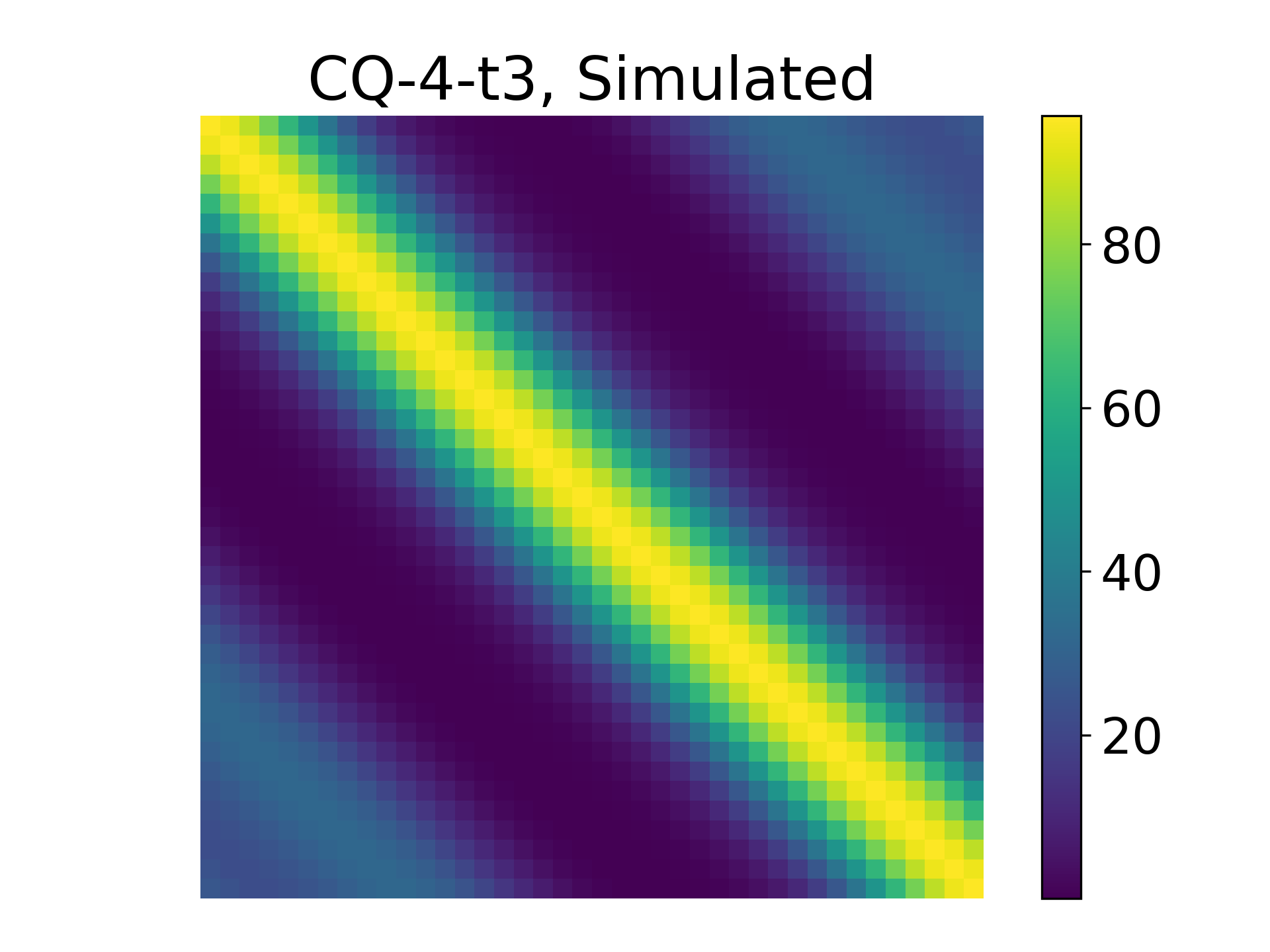}}
  \sidesubfloat[]{\includegraphics[width=0.33\columnwidth]{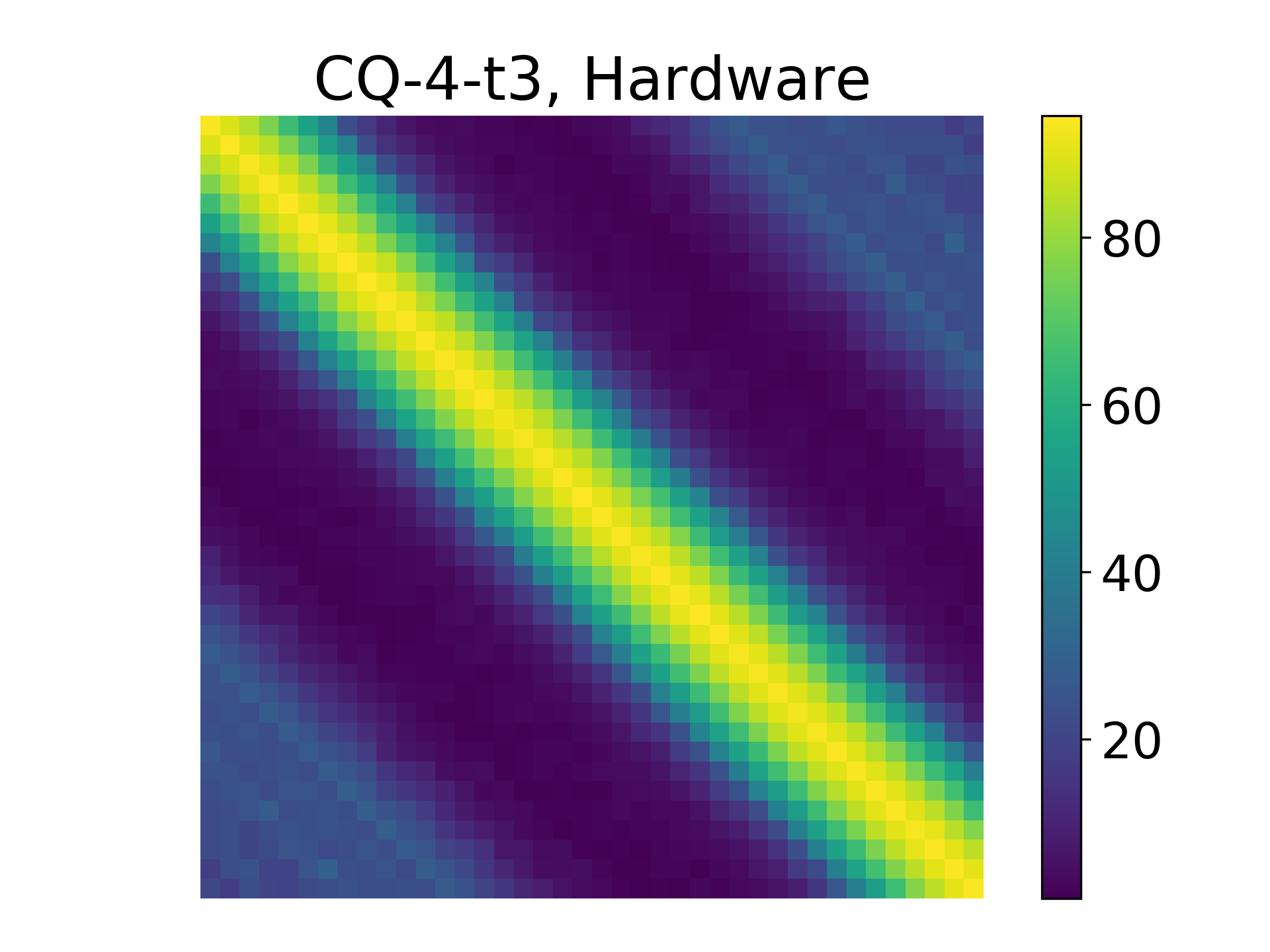}}
  \sidesubfloat[]{\includegraphics[width=0.33\columnwidth]{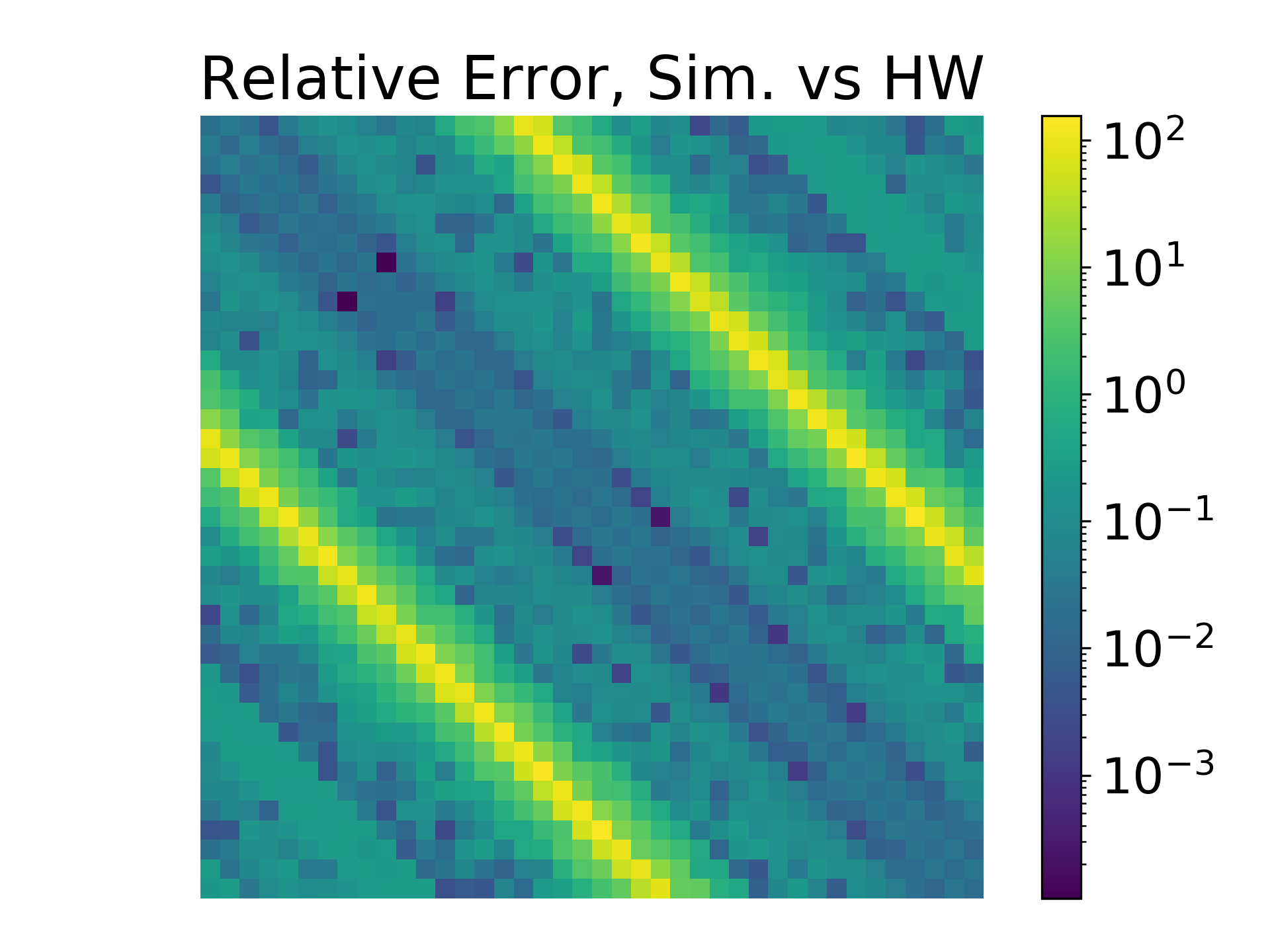}}
   \caption{Comparisons of the \texttt{CQ-4-t3} kernel, evaluated in simulation
     and on the IBM Boeblingen quantum computer. (a) The 
     \texttt{CQ-4-t3} Gram matrix from simulation. (b) The \texttt{CQ-4-t3} Gram
     matrix from hardware. (c) The relative error between the simulated and
     hardware Gram matrices.} \label{sim_vs_hw} 
\end{figure}

First, we show the results of evaluating the \texttt{CQ-4-t3} kernel directly on
the hardware. Note that this is different than the \texttt{CQ-4-t3-HW} kernel,
which includes an additional model discrepancy term (see main text).
Figure~\ref{sim_vs_hw}(a) shows the Gram matrix (i.e., the kernel matrix) of the
optimized \texttt{CQ-4-t3} kernel evaluated in simulation for the training
points used in the main text for the regression of $f(x)=x\sin(x)$. Since we
used a uniformly spaced grid of forty points, clear structure can be seen in the
Gram matrix. Points that are close to each other (near the diagonal) are much
more similar, represented by larger kernel values. There is a valley where
points are considered very distant from each other, according to this kernel.
Figure~\ref{sim_vs_hw}(b) shows the same kernel, with the same parameters and
training points, evaluated on the IBM Boeblingen quantum computer. The 
structure is very evident, and, upon first glance, one might consider the two
Gram matrices to approximately equal. However, as show in
Fig.~\ref{sim_vs_hw}(c), the relative error between the two kernels can be very
large. As explained in the main text, this is primarily due to the quantum
computer not being able to accurately compute the kernel value for points that
are very far away in the kernel space. This is represented by the large relative
error, approaching differences up to a factor of 100. The simulated kernel has a
minimum value of 0.016, while the kernel evaluated on the hardware has a minimum
value of 1.92 for the same pair of points. This leads us to develop a new
kernel, specifically for the hardware, which includes a model discrepancy term
to account for the hardware errors.

\subsection{Gaussian Process Prior Model for Quantum Hardware Experiments}

In this subsection, we formally write down the GP prior model for our quantum hardware 
experiments, and provide some additional context for results given in the main body
of the document. 
We restate our formulation of Eq.~(4) of the main text, this time explicitly including the
model discrepancy term.
We utilize a $\mathcal{GP}(\mathbf{0}, k(\cdot, \cdot))$ prior for $n$ possibly multi-variate inputs $X = \{\mathbf{x}_1, \dots,
\mathbf{x}_n\}$ and $n$ scalar outputs $\mathbf{y}$ given by
\begin{align}
  \mathbf{y} &= \mathbf{f} + \boldsymbol{\epsilon} + \mathbf{d}\\
  \mathbf{f} &\sim \mathcal{N}(\mathbf{0}, K_{X,X}) \\
  \boldsymbol{\epsilon}_i &\sim \mathcal{N}(0, \sigma_i^2) \\
  \mathbf{d}_i &\sim \mathcal{N}(0, \sigma_d^2).
\end{align}
Here $K_{X,X}$ is still an $n \times n$ matrix whose $(i, j)$th element is 
$k(\mathbf{x}_i, \mathbf{x}_j)$ and $\boldsymbol{\epsilon}$ is heteroscedastic 
measurement noise with prior variance vector 
$\mathbf{\sigma}^2 = [\sigma^2_1, \dots, \sigma^2_n]$.
$\mathbf{d}$, however, is a homoscedastic model discrepancy term that incorporates 
unmodeled modeling error arising from measuring the quantum hardware.  
We assume that each $\mathbf{d}_i$ is an unbiased Gaussian with unknown prior 
variance $\sigma_d^2$.  
This model precipitates the following joint distribution of $\mathbf{y}$ and the 
response $\mathbf{f}_*$ of $n_*$ unseen data $X_*$:
\begin{equation}
  \begin{bmatrix} \mathbf{y} \\ \mathbf{f}_*
  \end{bmatrix}
  = \mathcal{N} \left ( 0,
  \begin{bmatrix}
    K_{X,X} + \Sigma_d & K_{X, X_*} \\
    K_{X_*, X} & K_{X_*, X_*}
  \end{bmatrix} 
      \right ).
\end{equation}
Here, $\Sigma_d$ is a diagonal matrix whose $(i, i)$th element is 
$(\sigma^2_i + \sigma^2_d)$, incorporated the hardware-induced model discrepancy 
in the same form as the original GP prior presented in the main document.
This formulation gives us the similar posterior of the response 
$\mathbf{f}_* \sim \mathcal{N}(\bar{\mathbf{f}}_*, C)$, where
\begin{align}
  \bar{\mathbf{f}}_* &\equiv K_{X_*, X} J^{-1} \mathbf{y} \\ 
  C &\equiv K_{X_*,X_*} - K_{X_*,X} J^{-1} K_{X, X_*} \\
  J &\equiv K_{X,X} + \Sigma_d. 
\end{align}

\subsection{\texttt{CQ-4-t3-HW} Kernel}
\begin{figure}
  \centering
  \includegraphics[width=0.47\columnwidth]{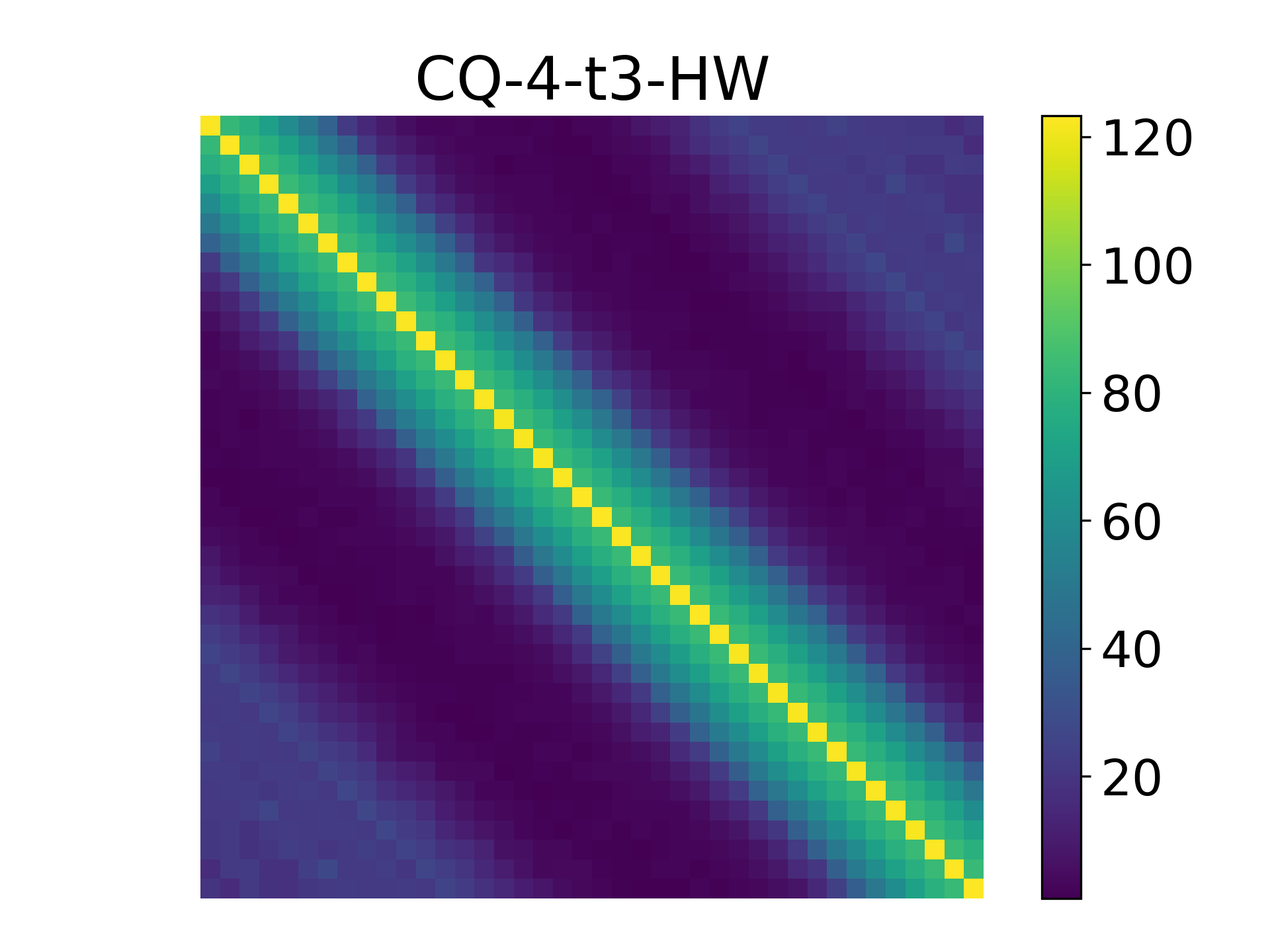}
  \caption{Gram matrix for the optimized \texttt{CQ-4-t3-HW} kernel, which
    includes the model discrepancy term.} \label{cq_4_t3_hw} 
\end{figure}

To optimize the hyperparameters for the \texttt{CQ-4-t3-HW} kernel, we first take
the values of the Gram matrix for the \texttt{CQ-4-t3} with the scaling
hyperparameter removed. This effectively fixes the $c_i$ parameter (to the value
2.225), as defined
above, but allows us to optimize the scaling hyperparameter ($s$ in the main
text) and the strength of
the model discrepancy term ($\sigma_d$ above) without making further calls to
the quantum device. Figure~\ref{cq_4_t3_hw} shows the resulting Gram matrix. Due
to the inclusion of the model discrepancy term, the diagonal has a far greater
value than the surrounding points.The general structure is still maintained. The
optimal parameters were as follows: $s$=7.469e+01, $\sigma_d$ = 3.074e+01. The
bounds for $s$ were [1e-2,1e2] and for $\sigma_d$ were [1e-3,1e3].

\section{Analytic Formula of Squeezed Kernel}
In this section, we derive the analytic formula for the displaced, two-mode
squeezed state that makes up the squeezed kernel used in the main text. The
feature map for the squeezed kernel in three-dimensions is
\begin{equation}\label{squeezed_fm}
  |\phi(x_1,x_2,x_3)\rangle = \prod_{\{i,j\}} D_i(\alpha_i) D_j(\alpha_j) S_{ij}(\gamma_{ij}) |0,0\rangle,
\end{equation}
where $D_i$ is the displacement operator on system $i$, $\alpha_i =
\frac{x_i}{\sqrt{2}c_i}$, $S_{i,j}$ is the two-mode
squeezing operator on systems $i$ and $j$ (see below), and the product is only taken over
$\{i,j\}$ pairs such the $i>j$. The corresponding kernel is defined as
\begin{equation}\label{squeezed_kernel}
  K(\mathbf{x},\mathbf{x'}) = \prod_{\{i,j\}} k\Big((x_i,x_j),(x'_i,x'_j)\Big) = s \prod_{\{i,j\}} \Big|\langle 0,0 |S^\dagger_{ij}(\gamma'_{ij}) D^\dagger_j(\alpha_j') D^\dagger_i(\alpha'_i) D_i(\alpha_i) D_j(\alpha_j) S_{ij}(\gamma_{ij}) |0,0 \rangle\Big|.
\end{equation}
Note that this kernel does not include the square that the coherent kernel of
the main text includes; this is because, as we will see, the squeezed kernel in
the limit of zero squeezing double counts each data dimension.

We define the 
two-mode squeezed state with squeezing parameter $\gamma_{i,j}$ as~\cite{schumaker1985new}
\begin{equation}\label{tmsv}
  \exp(\gamma_{ij} (b_i b_j - b_i^\dagger b_j^\dagger)) |0,0\rangle = \frac{1}{\cosh(\gamma_{ij})} \sum_{n=0}^\infty (-e^{i\Phi} \tanh(\gamma_{ij}))^n |n,n\rangle,
\end{equation}
where $\Phi$ is a phase. For now, we will focus on just one pair of ${i,j}$
within the squeezed feature map, eq.~\eqref{squeezed_fm}. The
displacement operators $D_i$ and $D_j$ can be brought into the infinite sum of the
two-mode squeezed state, eq.~\eqref{tmsv}
\begin{equation}
  \frac{1}{\cosh(\gamma_{ij})} \sum_{n=0}^\infty (-e^{i\Phi} \tanh(\gamma_{ij}))^n D_i(\alpha_i)D_j(\alpha_j)|n,n\rangle = \frac{1}{\cosh(\gamma_{ij})} \sum_{n=0}^\infty (-e^{i\Phi} \tanh(\gamma_{ij}))^n |(\alpha_i,n),(\alpha_j,n)\rangle,
\end{equation}
where $|(\alpha_i,n),(\alpha_j,n)\rangle$
represent displaced number states in each dimension.
We define the displaced number state as the displacement operator acting on a
number state, $|\alpha,n\rangle = D(\alpha)|n\rangle$ and note that the overlap
between two displaced number states has an analytic form~\cite{wunsche1991displaced}
\begin{equation}
  \langle m,\alpha' | n,\alpha\rangle = \langle \alpha' | \alpha \rangle \sqrt{\frac{m!}{n!}} (\alpha' - \alpha)^{n-m} L_m^{n-m}\big((\alpha'-\alpha)^2\big),
\end{equation}
where $L_m^{n-m}$ is the associated Laguerre polynomial. With the overlap
of displaced number states, we can now define the squeezed kernel for dimensions
$i,j$ as
\begin{equation}\label{sq_kernel_analytic}
  k\Big((x_i,x_j),(x'_i,x'_j)\Big) = \sum_{n=0,m=0}^\infty |\langle \alpha_i | \alpha_i' \rangle \langle \alpha_j | \alpha_j' \rangle  \tanh^n(\gamma_{i,j}) \tanh^m(\gamma'_{i,j}) L_m^{n-m}((\alpha'_i-\alpha_i)^2) L_m^{n-m}((\alpha_j'-\alpha_j)^2)|.
\end{equation}
Though the squeezed kernel has an analytic form, this infinite sum of associated
Laguerre polynomials does not have a clear analytic form. Furthermore, in the
$N$-mode squeezed case, or when squeezing is applied to multiple pairs of
dimensions together, it is not clear that an analytic form can be derived. We
approximate eq.~\eqref{sq_kernel_analytic} by truncating the sum to $n=8, m=8$,
which we found to be converged for the parameter ranges we used in the paper.

It is instructive to look at the limit of the analytic form of the squeezed
kernel, eq.~\eqref{sq_kernel_analytic}, as it approaches 0 squeezing
($\gamma_{i,j}=0$). Because 
$\tanh(0)=0$, the only term in the infinite sum that survives is the $n=0,m=0$
term, since $\tanh(0)^0 = 1$. The associated Laguerre polynomials for
$n=0,m=0$ are simply 1. As such, when there is zero squeezing, the squeezed
kernel simply becomes the coherent kernel. However, with a non-zero amount of
squeezing, the squeezed kernel is generating entanglement between data
dimensions and accessing parts of the Hilbert space which the coherent kernel
(and, therefore, the squared exponential kernel), cannot reach. Though there are
ways of making correlated squared exponential kernels, they do not have
the same form as squeezed kernel we have derived here. It uses the unique
entanglement properties of the underlying quantum representation; as a larger
number of modes are squeezed, the result accesses more and more of the
underlying Hilbert space. Truncated squeezed kernels could also be defined, by
using the truncated raising and lowering operators defined in the main text and above.

Though we have defined the analytic form of the squeezed kernel, we have not yet
defined the form of the squeezing parameter, $\gamma_{ij}$. The definition of
the displacement parameters, $\alpha_i$, were defined to be consistent with the
squared exponential kernel. Given that the squeezing parameter has no clear
classical analog, we are free to define it as we choose. One choice, inspired by
the definition of the displacement parameters, is to use the product of the data
points multiplied by a hyperparameter $d_{ij}$, $\gamma_{ij} = x_i x_j d_{ij}$.

\section{Additional Results for Dynamics Regression}
\begin{figure}
  \centering
  \sidesubfloat[]{\includegraphics[width=0.47\columnwidth]{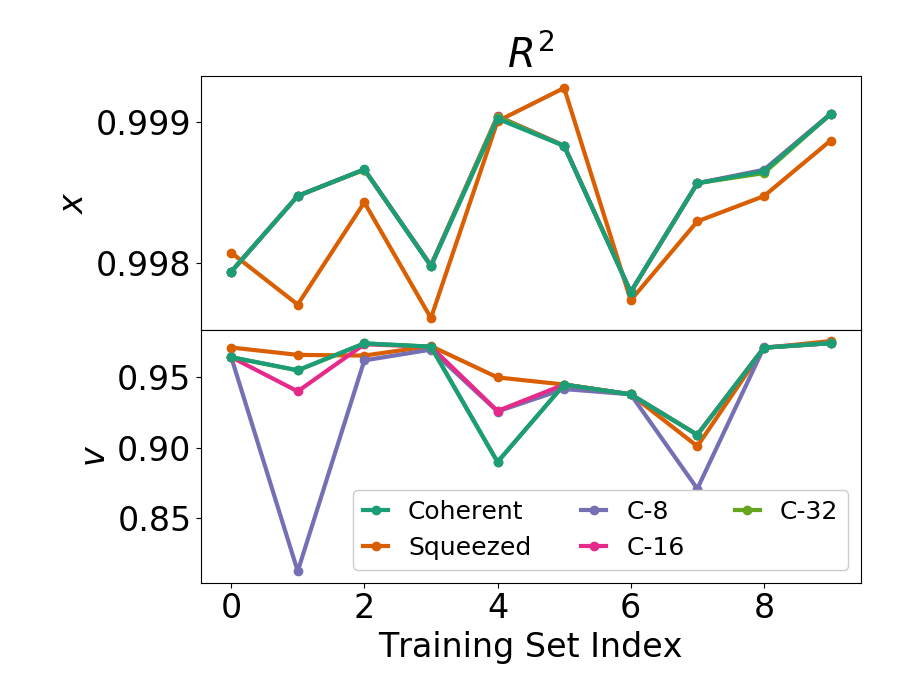}}
  \sidesubfloat[]{\includegraphics[width=0.47\columnwidth]{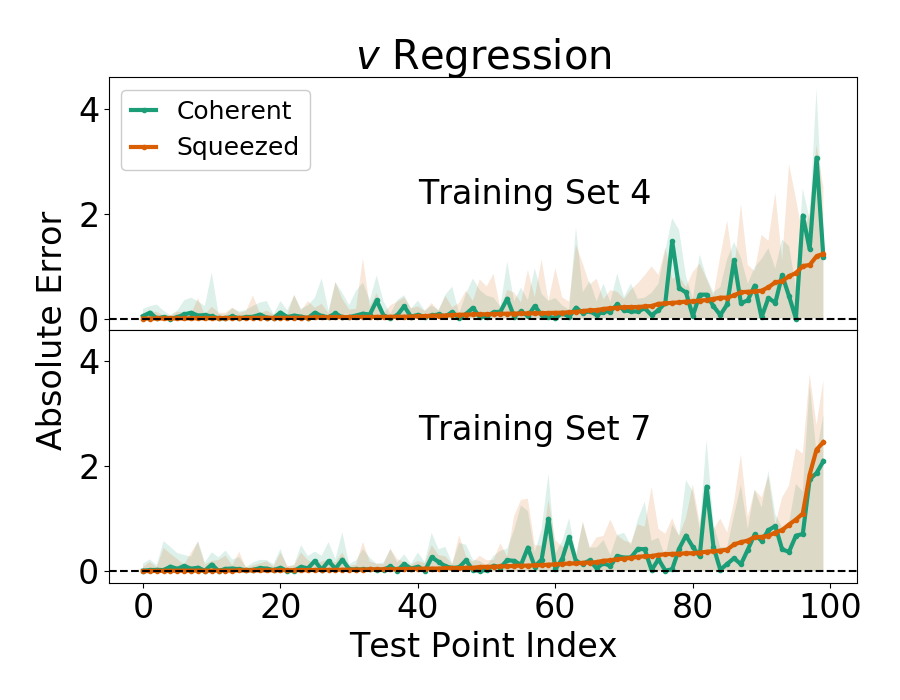}}
  \caption{Example of applying quantum kernels to dynamics regression.
    (a) The position, $x$, is regressed very well by all kernels. The velocity,
    $v$, needs a truncated Hilbert space of size 32 before the
    finite-dimensional coherent state regresses as well as the full coherent
    state. The addition of non-classical resources, through two-mode squeezing,
    allows the squeezed kernel to regress better than the coherent kernel. (b)
    Comparison of velocity regression using the squeezed and coherent kernels on
    100 random test points for two training sets. } \label{dynamics_regression} 
\end{figure}

In this section, we provide additional results and tables of hyperparameters for
the dynamics regression discussed in the main text.
Figure~\ref{dynamics_regression}(a) shows the $R^2$ values for each of the ten
training sets. In the regression of the position, $x$, all truncations perform
the same, while the squeezed kernel performs marginally better for some training
sets, but marginally worse for many others. The regression of the velocity, $v$,
is a harder regression task, as shown by the significantly smaller $R^2$
values. Here, the squeezed kernel performs better than the coherent kernel in
many of the training sets, and significantly better in training set four.
Figure~\ref{dynamics_regression}(b) shows the absolute errors in the prediction of the
velocity, $v$, in two training sets: four, where the squeezed kernel
vastly out performs the coherent kernel and seven, where the coherent kernel
marginally outperforms the squeezed kernel. In training set four, it is clear
that most of the points are regressed well by both the coherent and
squeezed kernels, but there are few outliers which the coherent kernel predicts
much worse than even the worst prediction of the squeezed kernel. In training
set seven, both the coherent and squeezed states perform generally the same,
with each having points that are better and worse.

\clearpage

\subsection{Optimal Hyperparameters for Dynamics Regression}
\setlength{\tabcolsep}{1em}

\begin{table*}[!htb]
  \begin{tabular}{ c| c| c| c| c| }
    \hline
    Training Set & $s$ & $c_x$ & $c_v$ & $c_a$ \\
    \hline
    0 & 3.709e-01 & 7.207e-01 & 3.432e+00 & 2.000e+01 \\
    1 & 3.738e-01 & 6.646e-01 & 3.360e+00 & 2.000e+01 \\
    2 & 3.714e-01 & 6.836e-01 & 2.919e+00 & 2.000e+01 \\
    3 & 4.380e-01 & 7.121e-01 & 3.122e+00 & 2.000e+01 \\
    4 & 3.716e-01 & 6.608e-01 & 2.984e+00 & 1.794e+01 \\
    5 & 3.994e-01 & 6.923e-01 & 3.061e+00 & 2.000e+01 \\
    6 & 3.269e-01 & 6.839e-01 & 3.143e+00 & 2.000e+01 \\
    7 & 4.277e-01 & 7.247e-01 & 3.092e+00 & 2.000e+01 \\
    8 & 3.689e-01 & 7.156e-01 & 3.106e+00 & 2.000e+01 \\
    9 & 4.027e-01 & 7.050e-01 & 2.988e+00 & 2.000e+01 \\
  \end{tabular}
  \caption{Optimal hyperparameters for $x$ regression using the coherent
    kernel. The $s$ bounds were [1e-3,1e3] and the $c_i$ bounds were [1e-3,20].}
\end{table*}

\begin{table*}[!htb]
  \begin{tabular}{ c| c| c| c| c| }
    \hline
    Training Set & $s$ & $c_x$ & $c_v$ & $c_a$ \\
    \hline
    0 & 2.094e+00 & 3.893e-01 & 7.742e-01 & 8.488e+00 \\
    1 & 2.613e+00 & 1.405e-01 & 1.426e+00 & 3.040e+00 \\
    2 & 1.705e+00 & 2.305e-01 & 1.347e+00 & 6.378e+00 \\
    3 & 2.944e+00 & 3.137e-01 & 1.473e+00 & 9.742e+00 \\
    4 & 2.510e+00 & 5.187e-01 & 3.305e-01 & 9.321e+00 \\
    5 & 1.503e+00 & 3.096e-01 & 1.121e+00 & 2.102e+00 \\
    6 & 2.453e+00 & 3.814e-01 & 1.506e+00 & 2.528e+00 \\
    7 & 2.390e+00 & 4.250e-01 & 3.888e-01 & 9.811e+00 \\
    8 & 2.676e+00 & 3.908e-01 & 7.673e-01 & 9.559e+00 \\
    9 & 3.312e+00 & 4.367e-01 & 1.608e+00 & 7.734e+00 \\
  \end{tabular}
  \caption{Optimal hyperparameters for $v$ regression using the coherent
    kernel. The $s$ bounds were [1e-3,1e3] and the $c_i$ bounds were [1e-3,20].}
\end{table*}

\begin{table*}[!htb]
  \begin{tabular}{ c| c| c| c| c| }
    \hline
    Training Set & $s$ & $c_x$ & $c_v$ & $c_a$ \\
    \hline
    0 & 3.708e-01 & 7.208e-01 & 3.434e+00 & 2.000e+01 \\
    1 & 3.771e-01 & 6.650e-01 & 3.371e+00 & 2.000e+01 \\
    2 & 3.721e-01 & 6.834e-01 & 2.926e+00 & 2.000e+01 \\
    3 & 4.368e-01 & 7.114e-01 & 3.124e+00 & 2.000e+01 \\
    4 & 3.589e-01 & 6.520e-01 & 2.945e+00 & 1.742e+01 \\
    5 & 3.983e-01 & 6.929e-01 & 3.053e+00 & 2.000e+01 \\
    6 & 3.270e-01 & 6.838e-01 & 3.142e+00 & 2.000e+01 \\
    7 & 4.277e-01 & 7.246e-01 & 3.093e+00 & 2.000e+01 \\
    8 & 3.622e-01 & 7.121e-01 & 3.069e+00 & 2.000e+01 \\
    9 & 4.056e-01 & 7.059e-01 & 2.994e+00 & 2.000e+01 \\
  \end{tabular}
  \caption{Optimal hyperparameters for $x$ regression using the \texttt{C-8}
    kernel. The $s$ bounds were [1e-3,1e3] and the $c_i$ bounds were [1e-3,20].}
\end{table*}
    
\begin{table*}[!htb]
  \begin{tabular}{ c| c| c| c| c| }
    \hline
    Training Set & $s$ & $c_x$ & $c_v$ & $c_a$ \\
    \hline
    0 & 2.012e+00 & 3.872e-01 & 7.620e-01 & 8.531e+00 \\
    1 & 8.837e+01 & 1.804e-01 & 2.131e+00 & 5.232e+00 \\
    2 & 1.883e+00 & 2.998e-01 & 1.270e+00 & 4.869e+00 \\
    3 & 3.474e+00 & 3.393e-01 & 1.434e+00 & 9.622e+00 \\
    4 & 1.712e+00 & 3.261e-01 & 9.069e-01 & 2.852e+00 \\
    5 & 1.427e+00 & 3.072e-01 & 1.114e+00 & 2.076e+00 \\
    6 & 2.443e+00 & 3.815e-01 & 1.498e+00 & 2.534e+00 \\
    7 & 1.854e+00 & 3.993e-01 & 6.510e-01 & 3.026e+00 \\
    8 & 2.786e+00 & 3.957e-01 & 7.697e-01 & 9.789e+00 \\
    9 & 3.431e+00 & 4.372e-01 & 1.615e+00 & 7.871e+00 \\
  \end{tabular}
  \caption{Optimal hyperparameters for $v$ regression using the \texttt{C-8}
    kernel. The $s$ bounds were [1e-3,1e3] and the $c_i$ bounds were [1e-3,20].}
\end{table*}

\begin{table*}[!htb]
  \begin{tabular}{ c| c| c| c| c| }
    \hline
    Training Set & $s$ & $c_x$ & $c_v$ & $c_a$ \\
    \hline
    0 & 3.710e-01 & 7.206e-01 & 3.432e+00 & 2.000e+01 \\
    1 & 3.757e-01 & 6.643e-01 & 3.360e+00 & 2.000e+01 \\
    2 & 3.726e-01 & 6.836e-01 & 2.926e+00 & 2.000e+01 \\
    3 & 4.354e-01 & 7.114e-01 & 3.117e+00 & 2.000e+01 \\
    4 & 3.464e-01 & 6.534e-01 & 2.890e+00 & 1.700e+01 \\
    5 & 3.986e-01 & 6.921e-01 & 3.055e+00 & 2.000e+01 \\
    6 & 3.270e-01 & 6.839e-01 & 3.142e+00 & 2.000e+01 \\
    7 & 4.276e-01 & 7.245e-01 & 3.093e+00 & 2.000e+01 \\
    8 & 3.713e-01 & 7.149e-01 & 3.105e+00 & 2.000e+01 \\
    9 & 4.029e-01 & 7.051e-01 & 2.987e+00 & 2.000e+01 \\
  \end{tabular}
  \caption{Optimal hyperparameters for $x$ regression using the \texttt{C-16}
    kernel. The $s$ bounds were [1e-3,1e3] and the $c_i$ bounds were [1e-3,20].}
\end{table*}

\begin{table*}[!htb]
  \begin{tabular}{ c| c| c| c| c| }
    \hline
    Training Set & $s$ & $c_x$ & $c_v$ & $c_a$ \\
    \hline
    0 & 2.064e+00 & 3.885e-01 & 7.730e-01 & 8.450e+00 \\
    1 & 3.670e+00 & 1.923e-01 & 1.323e+00 & 2.725e+00 \\
    2 & 1.701e+00 & 2.289e-01 & 1.349e+00 & 6.411e+00 \\
    3 & 3.035e+00 & 3.138e-01 & 1.473e+00 & 9.922e+00 \\
    4 & 1.634e+00 & 3.195e-01 & 9.065e-01 & 2.843e+00 \\
    5 & 1.504e+00 & 3.097e-01 & 1.122e+00 & 2.102e+00 \\
    6 & 2.451e+00 & 3.812e-01 & 1.507e+00 & 2.528e+00 \\
    7 & 3.002e+00 & 4.296e-01 & 4.160e-01 & 1.060e+01 \\
    8 & 2.691e+00 & 3.910e-01 & 7.689e-01 & 9.660e+00 \\
    9 & 3.415e+00 & 4.370e-01 & 1.615e+00 & 7.856e+00 \\
  \end{tabular}
  \caption{Optimal hyperparameters for $v$ regression using the \texttt{C-16}
    kernel. The $s$ bounds were [1e-3,1e3] and the $c_i$ bounds were [1e-3,20].}
\end{table*}

\begin{table*}[!htb]
  \begin{tabular}{ c| c| c| c| c| }
    \hline
    Training Set & $s$ & $c_x$ & $c_v$ & $c_a$ \\
    \hline
    0 & 3.709e-01 & 7.207e-01 & 3.431e+00 & 2.000e+01 \\
    1 & 3.745e-01 & 6.642e-01 & 3.362e+00 & 2.000e+01 \\
    2 & 3.723e-01 & 6.840e-01 & 2.926e+00 & 2.000e+01 \\
    3 & 4.364e-01 & 7.116e-01 & 3.120e+00 & 2.000e+01 \\
    4 & 3.599e-01 & 6.519e-01 & 2.950e+00 & 1.747e+01 \\
    5 & 3.987e-01 & 6.925e-01 & 3.058e+00 & 2.000e+01 \\
    6 & 3.269e-01 & 6.840e-01 & 3.142e+00 & 2.000e+01 \\
    7 & 4.279e-01 & 7.244e-01 & 3.095e+00 & 2.000e+01 \\
    8 & 3.763e-01 & 7.099e-01 & 3.180e+00 & 2.000e+01 \\
    9 & 4.033e-01 & 7.051e-01 & 2.989e+00 & 2.000e+01 \\
  \end{tabular}
  \caption{Optimal hyperparameters for $x$ regression using the \texttt{C-32}
    kernel. The $s$ bounds were [1e-3,1e3] and the $c_i$ bounds were [1e-3,20].}
\end{table*}

\begin{table*}[!htb]
  \begin{tabular}{ c| c| c| c| c| }
    \hline
    Training Set & $s$ & $c_x$ & $c_v$ & $c_a$ \\
    \hline
    0 & 2.096e+00 & 3.893e-01 & 7.745e-01 & 8.491e+00 \\
    1 & 2.598e+00 & 1.399e-01 & 1.426e+00 & 3.043e+00 \\
    2 & 1.705e+00 & 2.305e-01 & 1.347e+00 & 6.373e+00 \\
    3 & 3.040e+00 & 3.140e-01 & 1.472e+00 & 9.928e+00 \\
    4 & 2.505e+00 & 5.184e-01 & 3.305e-01 & 9.305e+00 \\
    5 & 1.503e+00 & 3.094e-01 & 1.123e+00 & 2.103e+00 \\
    6 & 2.455e+00 & 3.815e-01 & 1.507e+00 & 2.527e+00 \\
    7 & 2.379e+00 & 4.247e-01 & 3.885e-01 & 9.789e+00 \\
    8 & 2.691e+00 & 3.909e-01 & 7.690e-01 & 9.659e+00 \\
    9 & 3.398e+00 & 4.366e-01 & 1.615e+00 & 7.844e+00 \\
  \end{tabular}
  \caption{Optimal hyperparameters for $v$ regression using the \texttt{C-32}
    kernel. The $s$ bounds were [1e-3,1e3] and the $c_i$ bounds were [1e-3,20].}
\end{table*}

\begin{table*}[!htb]
  \begin{tabular}{ c| c| c| c| c| c| c| c }
    \hline
    Training Set & $s$ & $c_x$ & $c_v$ & $c_a$ & $\gamma_{xv}$ & $\gamma_{va}$ & $\gamma_{xa}$ \\
    \hline
    0 & 5.048e-01 & 8.110e-01 & 5.663e+00 & 2.000e+01 & 2.213e-01 & 2.171e-05 & 1.622e-04\\
    1 & 4.462e-01 & 7.857e-01 & 5.390e+00 & 2.000e+01 & 2.507e-01 & 3.996e-02 & 4.057e-03\\
    2 & 4.827e-01 & 8.284e-01 & 5.771e+00 & 2.000e+01 & 2.468e-01 & 7.876e-05 & 6.075e-05\\
    3 & 6.271e-01 & 8.332e-01 & 5.980e+00 & 2.000e+01 & 2.464e-01 & 2.159e-03 & 5.196e-03\\
    4 & 4.374e-01 & 7.716e-01 & 5.806e+00 & 2.000e+01 & 2.685e-01 & 6.170e-02 & 2.121e-03\\
    5 & 4.405e-01 & 7.653e-01 & 5.602e+00 & 2.000e+01 & 2.899e-01 & 2.476e-02 & 1.078e-05\\
    6 & 4.501e-01 & 7.873e-01 & 5.874e+00 & 2.000e+01 & 2.314e-01 & 1.667e-02 & 5.066e-03\\
    7 & 5.947e-01 & 8.226e-01 & 5.947e+00 & 2.000e+01 & 2.293e-01 & 3.600e-04 & 2.168e-19\\
    8 & 6.077e-01 & 8.394e-01 & 6.424e+00 & 2.000e+01 & 2.142e-01 & 1.742e-03 & 1.320e-04\\
    9 & 6.184e-01 & 8.372e-01 & 6.548e+00 & 2.000e+01 & 2.128e-01 & 2.329e-05 & 9.595e-04\\
  \end{tabular}
  \caption{Optimal hyperparameters for $x$ regression using the squeezed
    kernel. The $s$ bounds were [1e-3,1e3], the $c_i$ bounds were [1e-3,20], and
  the $\gamma_{ij}$ bounds were [0,19.999].}
\end{table*}

\begin{table*}[!htb]
  \begin{tabular}{ c |c |c |c |c |c |c| c }
    \hline
    Training Set & $s$ & $c_x$ & $c_v$ & $c_a$ & $\gamma_{xv}$ & $\gamma_{va}$ & $\gamma_{xa}$ \\
    \hline
    0 & 2.259e+00 & 3.884e-01 & 8.701e-01 & 1.149e+01 & 5.810e-04 & 1.430e-03 & 7.114e-02\\
    1 & 4.630e+00 & 1.906e-01 & 2.343e+00 & 1.195e+01 & 5.282e-04 & 4.927e-04 & 1.653e-01\\
    2 & 2.344e+00 & 3.185e-01 & 1.651e+00 & 1.089e+01 & 3.060e-04 & 5.369e-04 & 1.247e-01\\
    3 & 2.949e+00 & 3.134e-01 & 1.490e+00 & 1.040e+01 & 2.168e-19 & 6.097e-02 & 1.256e-04\\
    4 & 2.279e+00 & 3.149e-01 & 1.068e+00 & 1.091e+01 & 1.585e-05 & 3.703e-04 & 1.886e-01\\
    5 & 1.496e+00 & 3.092e-01 & 1.125e+00 & 2.098e+00 & 5.487e-05 & 3.621e-04 & 3.740e-04\\
    6 & 2.455e+00 & 3.816e-01 & 1.506e+00 & 2.527e+00 & 5.415e-06 & 1.073e-05 & 4.156e-04\\
    7 & 2.024e+00 & 4.417e-01 & 5.301e-01 & 1.217e+01 & 3.035e-03 & 1.661e-03 & 1.279e-01\\
    8 & 2.671e+00 & 3.898e-01 & 7.671e-01 & 9.641e+00 & 1.028e-03 & 3.455e-03 & 2.168e-19\\
    9 & 2.536e+00 & 4.993e-01 & 2.499e+00 & 7.991e+00 & 8.290e-01 & 7.193e-05 & 3.672e-04\\
  \end{tabular}
  \caption{Optimal hyperparameters for $v$ regression using the squeezed kernel.
    The $s$ bounds were [1e-3,1e3], the $c_i$ bounds were [1e-3,20], and
    the $\gamma_{ij}$ bounds were [0,19.999].}
\end{table*}

\clearpage

\section{Optimal Hyperparameters for Reinforcement Learning Results}

\begin{table*}[!htb]
  \begin{tabular}{ c |c |c |c }
    \hline
    Kernel & $s$ & $c_x$ & $c_v$ \\
    \hline
    Coherent & 4.753e+01 & 1.469e-01 & 4.587e-01 \\
    \texttt{C-8} & 2.295e+01 & 2.427e-01 & 1.008e-01 \\
    \texttt{C-16} & 4.882e+01 & 1.865e-01 & 4.726e-01 \\
  \end{tabular}
  \caption{Optimal hyperparameters for value regression in the reinforcement
    learning task.
    The $s$ bounds were [1e-3,1e2] and the $c_i$ bounds were [0.05,10].}
\end{table*}

\begin{table*}[!htb]
  \begin{tabular}{ c |c |c |c |c  }
    \hline
    Kernel & $s$ & $c_x$ & $c_v$ & $c_a$ \\
    \hline
    Coherent & 3.827e-01 & 6.599e-01 & 3.292e+00 & 1.984e+01 \\
    \texttt{C-8} & 3.516e-01 & 8.144e-01 & 3.121e+00 & 2.000e+01 \\
    \texttt{C-16} & 4.405e-01 & 7.561e-01 & 3.981e+00 & 2.000e+01 \\
  \end{tabular}
  \caption{Optimal hyperparameters for $x$ regression in the reinforcement
    learning task.
    The $s$ bounds were [1e-3,1e3] and the $c_i$ bounds were [1e-3,20].}
\end{table*}

\begin{table*}[!htb]
  \begin{tabular}{ c |c |c |c |c  }
    \hline
    Kernel & $s$ & $c_x$ & $c_v$ & $c_a$ \\
    \hline
    Coherent & 6.111e+00 & 3.710e-01 & 2.334e+00 & 1.656e+01 \\
    \texttt{C-8} & 1.969e+00 & 5.406e-01 & 1.675e+00 & 4.988e+00 \\
    \texttt{C-16} & 3.304e+00 & 4.852e-01 & 1.842e+00 & 7.301e+00 \\
  \end{tabular}
  \caption{Optimal hyperparameters for $v$ regression in the reinforcement
  learning task.
  The $s$ bounds were [1e-3,1e3] and the $c_i$ bounds were [1e-3,20].}
\end{table*}

\clearpage


%